\documentclass[twocolumn,iop,numberedappendix,appendixfloats]{openjournal}

\usepackage[table, dvipsnames,svgnames,x11names]{xcolor}
\usepackage[colorlinks=true, linktocpage, linkcolor={blue!60!black}, citecolor={blue!60!black}, urlcolor={blue!60!black}]{hyperref}
\usepackage{scrextend}
\usepackage{amsmath}
\usepackage{amssymb}
\usepackage{graphicx}
 \usepackage{booktabs}

\usepackage{txfonts}
\usepackage{enumitem}
\usepackage{subcaption}
\usepackage{multirow}
\usepackage{rotating}

\newcommand{\updated}[1]{#1}

\def\RoCS{1}
\def\ITA{2}
\def\LEIDEN{3}
\def\ESO{4}
\def\csicice{5}
\def\YALE{6}
\def\BrookhavenLab{7}

\begin{document} 
    \journalinfo{The Open Journal of Astrophysics}
    %\submitted{submitted; X, accepted; X}

    \title{Observing the Sun with the Atacama Large Aperture Submillimeter Telescope (AtLAST): Forecasting Full-disk Observations}

    \author{M.~Kirkaune\altaffilmark{\RoCS, \ITA}, 
            S.~Wedemeyer\altaffilmark{\RoCS, \ITA},
            J.~van Marrewijk\altaffilmark{\LEIDEN},
            T.~Mroczkowski\altaffilmark{\ESO, \csicice},
            and
            T.~W.~Morris\altaffilmark{\YALE, \BrookhavenLab}
            }
    
    % List of institutions
    
    \affiliation{\altaffilmark{\RoCS} Rosseland Centre for Solar Physics, University of Oslo, Postboks 1029 Blindern, N-0315 Oslo, Norway}

    \affiliation{\altaffilmark{\ITA} Institute of Theoretical Astrophysics, University of Oslo, Postboks 1029 Blindern, N-0315 Oslo, Norway}

    \affiliation{\altaffilmark{\LEIDEN} Leiden Observatory, Leiden University, P.O. Box 9513, 2300 RA Leiden, The Netherlands}

    \affiliation{\altaffilmark{\ESO} European Southern Observatory (ESO), Karl-Schwarzschild-Strasse 2, Garching 85748, Germany}

    \affiliation{\altaffilmark{\csicice} Institute of Space Sciences (ICE, CSIC), Carrer de Can Magrans, s/n, 08193 Cerdanyola del Vallès, Barcelona, Spain}

    \affiliation{\altaffilmark{\YALE} Department of Physics, Yale University, New Haven, CT 06511, USA}
    
    \affiliation{\altaffilmark{\BrookhavenLab} Brookhaven National Laboratory, Upton, NY 11973, USA}
    
\begin{abstract}
The Atacama Large Millimeter Array (ALMA) has revolutionised the field of solar millimetre astronomy with its high angular resolution and cadence. However, interferometers often have very limited fields of view (FOV), making targeted observations of highly dynamic phenomena such of flares challenging, as they are hard to predict. A large aperture single-dish millimetre telescope with a large FOV, such as the future Atacama Large Aperture Submillimeter Telescope (AtLAST), would prove useful in observing such phenomena, as one could scan the full solar disk on shorter timescales than current millimetre facilities. We aim to explore what FOVs, detector counts, and scan strategies are suitable for AtLAST to realistically push the required full-disk scan times below 1 minute, enabling regular observations of dynamic phenomena on the Sun. Utilising the \texttt{maria} code, we have been able to realistically simulate solar observations with AtLAST, and thoroughly explored how instrumental properties and scanning strategies affect the full-disk observations at the different frequency bands planned for AtLAST. As input images we used realistic solar disk maps, initially constructed from publicly available full disk maps from the Solar Dynamics Observatory (SDO), and thoroughly compared to ALMA total power maps. We find the double-circle scan pattern, currently employed at ALMA for full-disk mapping  to be an acceptable way of scanning the Sun with AtLAST as well. Using small to intermediately sized instruments (1000 - 50,000 detector elements), the estimated observational cadence would be less than 1 minute across AtLAST's frequency range with a reasonable pixel spacing. Using instruments with larger FOVs ($\gtrapprox 0.25^\circ$, equivalent to $\gtrapprox$ 1 R$_\odot$), we find a simple circular scan to be more efficient, achieving cadences on second time scales across the frequency range, but requiring more detector elements ($\gtrapprox$ 100,000). We find that a large FOV single-dish telescope such as AtLAST could provide the solar millimetre community with hitherto unachievable observations, namely full-disk observations at high cadence and adequate resolution. With cadences potentially down to seconds, such an instrument would be ideal in the study of quickly evolving phenomena, such as flares.
\end{abstract}

\section{Introduction}
Solar millimetre and submillimetre radiation mostly originates in the chromosphere, the atmospheric layer in-between the relatively cool photosphere and the hot corona \citep{2016SSRv..200....1W}. How the corona is heated to temperatures of the order $10^6$ K, three orders of magnitude hotter than the photosphere, has puzzled astronomers for decades and is known as the coronal heating problem. 
As all energy that is emitted from the Sun's photosphere has to pass through the chromosphere, this layer plays a role in the quest to solve the atmospheric heating problem. 
Imprints of heating phenomena might be observed in the chromosphere, thus constraining their contributions to heating the atmosphere. The chromosphere can be studied through spectroscopic measurements in the visible and ultra-violet (UV) bands \citep{2020A&A...641A.146R, 2025A&A...693A...8T}. However, these lines are formed under non-local thermodynamic equilibrium conditions, and are therefore hard to interpret \citep[see, e.g.][]{2016ApJ...830L..30D,2018Msngr.171...25B}. Millimetre and submillimetre radiation from the chromosphere has the advantage of being thermal continuum emission, therefore formed under local thermodynamic equilibrium, and consequently much easier to interpret \citep{2018Msngr.171...25B}. Furthermore, being within the Rayleigh-Jeans limit has the advantage that the observed intensity is linearly proportional to the temperature of the emitting plasma. 
%, (sub-)millimetre observations are a powerful tool for understanding the chromosphere and the dynamic phenomena it encompasses. 
Additionally, the formation height for this continuum flux has the advantage of increasing with wavelength \citep[see][and references therein]{2016SSRv..200....1W}, thus making multi-frequency solar observations in the mm-regime ideal for exploring how the chromospheric plasma temperature varies as a function of height over regions of the Sun with varying activity levels such as the Quiet Sun (QS), coronal holes, and Active Regions (ARs). This is important in trying to understand how energy is transported into and through the chromosphere, which plays a role in better understanding the atmospheric heating problem.

The range of solar science cases that can be explored in the millimetre regime is wide, spanning from the study of flares to the magnetic topology of the Sun and the solar-stellar connection \citep{2002AN....323..271B,2018Msngr.171...25B,2016SSRv..200....1W,2024ORE.....4..140W}. The large range of solar science cases AtLAST could contribute to also introduces different observational constraints AtLAST observations would have to meet to study each science case. The major solar science cases relevant for AtLAST, as well as the constraints each science case put on future AtLAST observations are introduced in Sect.~\ref{sec:sc_cases}.

The Atacama Large Millimeter Array \citep[ALMA,][]{2009IEEEP..97.1463W} has for almost a decade been one of the main drivers of solar millimetre astronomy \citep[see][and references therein]{2022FrASS...9.7368B}, with its superior spectral- and angular resolution and high cadence to other facilities observing in the same frequency range \citep{2018Msngr.171...25B,2022FrASS...9.7368B, 2016SSRv..200....1W}. Ongoing and future  improvements to instrumentation, data processing, and calibration methods will facilitate excellent solar observations in the mm-regime for decades to come.  

Its high resolution and excellent cadence makes ALMA a powerful tool in studying the small-scale dynamic phenomena in the chromosphere. However, the field of view of ALMA is limited, making observations of extended phenomena problematic. 
As solar flares are hard to predict, and observing campaigns with ALMA are planned weeks or months in advance, observing flares with ALMA is notoriously difficult, mostly reducing it to few coincidental detections  
\citep{2021ApJ...922..113S, 2023A&A...669A.156S}. Additionally, the small FOV makes ALMA unsuitable for full-disk mapping in its interferometric configuration, as the required time to cover the disk is high, especially when one considers the highly dynamic nature of the solar chromosphere. For full-disk mapping, ALMA utilises a single total power (TP) antenna to scan the full disk on time scales of $\sim$10~minutes \citep{2017SoPh..292...88W}, but at much lower resolution compared to its interferometric observations. The resulting full-disk TP maps, although invaluable in acting as the zeroth-baseline to correctly offset ALMA's interferometric observations, are of inadequate resolution and are too long in-between to study the evolution of several phenomena in the solar chromosphere observable in the millimetre regime, such as small flares and bursts. For these phenomena to be meaningfully studied, a higher resolution instrument with a larger field of view is required.

The Atacama Large Aperture Submillimeter Telescope (AtLAST), a single-dish telescope currently under development will combine the high angular resolution of a 50-metre aperture with a FOV of up to 2$^\circ$ \citep{2024arXiv240218645M}, enabling fast scanning of the millimetre universe including the Sun \citep{2024ORE.....4..140W}. With six instrument bays, building either a dedicated solar instrument, or a multi-purpose instrument is a possibility  that could provide the scientific community with high quality solar millimetre data. The combination of a large aperture, a multi-pixel instrument with a large FOV, and fast-scanning capabilities could produce high-resolution and high cadence observations of the full solar disk, allowing the study of solar phenomena that are difficult to study with ALMA, such as flares and large-scale active regions. The properties of AtLAST itself are now clearly defined in \citet{2024arXiv240218645M}, but the properties of a possible solar- or multi-purpose instrument are still undecided.

To aid in the decision of favourable instrumental properties for a possible future solar-capable instrument, the capabilities of different hypothetical instruments for AtLAST are here tested, with an emphasis on the capability for high-cadence full-disk observations. To achieve this, the all-purpose single-dish telescope simulator \texttt{maria}\footnote{\href{https://thomaswmorris.com/maria/}{https://thomaswmorris.com/maria/}} \citep{2024OJAp....7E.118V} is used.

Section~\ref{sec:data_methods} provides a brief overview of the \texttt{maria} simulator, and how it is used to simulate AtLAST observations of the Sun. In Sect.~\ref{sec:results}, these synthetic observations are used to test different instrumental setups and properties, ultimately to explore the achievable cadence of full-disk observations. 
A discussion of the findings is given in Sect.~\ref{sec:discussion}. The suitability of various instruments are here discussed in detail, giving some insight into the kind of instrument one should consider for installation at AtLAST in the future. A brief conclusion of our work is found in Sect.~\ref{sec:conclusion}

\section{Solar Science Cases for AtLAST} \label{sec:sc_cases}

The wide range of possible solar science cases that could be studied with AtLAST are described in detail in \citet{2024ORE.....4..140W}. Here, a brief summary of the various solar science cases and their observational requirements is provided. %We refer the readers to \citet{2024ORE.....4..140W} for more details.

\subsection{Thermal structure and heating of the solar atmosphere} \label{sec:sc_heating}
The corona, the outermost layer of the solar atmosphere is orders of magnitude hotter than the solar surface, the photosphere \citep{2015RSPTA.37340269D}, which is known as the atmospheric- or coronal heating problem. The coronal heating problem is one of the greatest unanswered questions in (solar) astrophysics, and decades of research has provided many processes capable of heating the atmosphere, although the extent with which each process contributes is unknown \citep{2024ORE.....4..140W}. 

AtLAST could contribute to the study of the heating of the solar atmosphere, both by observing possible candidates for heating mechanisms, e.g. in connection with flares (see Sect.~\ref{sec:sc_flares}), but also through studying the thermal structure of the solar atmosphere. The millimetre regime is suited for this, as the (sub-)millimetre radiation is formed under local thermodynamic equilibrium and within the Rayleigh-Jeans limit. The observed brightness temperature of the (sub-)mm radiation is therefore linearly dependent on the local plasma temperature, thus AtLAST could function as a linear thermometer, measuring the local plasma temperature in the solar atmosphere. By observing, preferably simultaneously, in different frequency bands, AtLAST would probe \updated{the temperature at} different heights in the atmosphere, thus providing valuable insight into the thermal structure of the solar atmosphere. \updated{This would better constrain the heights at which heating occurs, and therefore yielding important information regarding what mechanisms are likely to cause said heating}.

\subsection{Solar flares} \label{sec:sc_flares}
Solar flares are observationally defined to be brightenings in the solar atmosphere at time scales of minutes to hours, and are observed all across the electromagnetic spectrum \citep{2017LRSP...14....2B}. These brightenings are due to heating of the local plasma, usually attributed to the sudden release of magnetic energy by magnetic reconnection, and can occur in regions where the magnetic field is concentrated \citep{1997ApJ...488..499K, 1998SoPh..182..349B, 1998A&A...336.1039B}, but most often occur in active regions where the magnetic field is strong and highly complex \citep{2006A&A...451..319R}. The largest flares are also often accompanied by coronal mass ejections (CME), with implications for space weather, potentially affecting Earth\citep{2017LRSP...14....2B, 2024ORE.....4..140W}.

The study of solar flares with AtLAST poses some observational constraints that should be taken into account when considering a solar-capable instrument for the facility. Firstly, a large FOV would aid in increasing the probability of serendipitously detecting flares and other transient brightenings, the larger the FOV the larger the probability. Additionally, a sub-minute temporal cadence would be beneficial to sufficiently resolve the evolution of the shorter lived flares, preferably $< 10$ seconds to sufficiently capture the flare's evolution. There is some uncertainty regarding the time scales of the smallest flares, but microflares have been observed in the millimetre regime with lifetimes on the order of minutes \citep{2021ApJ...922..113S}. It is however likely that much shorter lived flares are present and detectable in solar millimetre radiation. 

\subsection{Solar prominences} \label{sec:prominences}
Solar prominences are large and highly extended strands of relatively cool plasma stretching all the way from the cool photosphere into the much hotter (surrounding) solar corona \citep[see,][for a detailed overview of solar prominences]{2015ASSL..415.....V}. The complex magnetic field of the prominence both supports it against gravity, and insulates its cool plasma from the much hotter surrounding environment. Some prominences become unstable over time, and erupting prominences play an important role as a driver of space weather. For meaningful observations of prominences with AtLAST a large FOV is crucial because of the large spatial scales of prominences. Solar prominences can be hundreds of Mm long, even exceeding one solar radius in length in some cases \citep{2010SSRv..151..333M}. The angular scale of one solar radius is roughly $\sim 0.25^\circ$, meaning that an instantaneous FOV of at least $\sim 0.25^\circ$ would be required to instantaneously observe the largest prominences. However, it is likely that a smaller FOV instrument could be used to scan large prominences instead, trading temporal resolution for using a smaller instrument. Additionally, the possibility of a multi-frequency instrument could prove instrumental in the study of prominences' thermal properties. Finally, studying the temporal evolution of prominences through daily multi-band maps over long periods of time would be an invaluable contribution towards our understanding of the evolution of solar prominences.

\subsection{The solar cycle}
The solar activity level varies periodically over a period of $\sim 11$ years between each solar maximum. AtLAST could study the solar cycle by providing insight into the thermal structure of the solar atmosphere in regions of varying activity levels, such as the quiet Sun and active regions. However, the study of the solar cycle does not introduce any particular observational constraints, as a high cadence will not be required to study the slowly evolving solar activity cycle. Instead, this introduces a requirement to the operation of the facility itself, as full-disk maps should be created with a set interval, for example one per day. Having a time stream of such maps over many years would contribute greatly to the study of the Sun's varying activity, and is something AtLAST could provide to the community. 

\newpage
\section{Data and methods} \label{sec:data_methods}

\subsection{The \texttt{maria} simulator}
To investigate possible solar instruments for AtLAST and their ability to meet the observational requirements set by the solar science cases discussed in Sect.~\ref{sec:sc_cases}, a method of simulating realistic single-dish millimetre observations is required. We used the \texttt{maria} code \citep{ 2022PhRvD.105d2004M, 2024OJAp....7E.118V}, a new simulator for forecasting single-dish \mbox{(sub-)mm} observations. The \texttt{maria} code allows for simulating a large range of highly customisable instruments, and to test their suitability for observations by employing different scan strategies and different seeing conditions. The generated, evolving atmosphere in \texttt{maria} is both location- and time-specific, and varies greatly between the different site selections available \citep[see][for an overview]{2022PhRvD.105d2004M}. 

\begin{figure*}[]
    \vspace{-1mm}
    \centering
\includegraphics[width=1\linewidth]{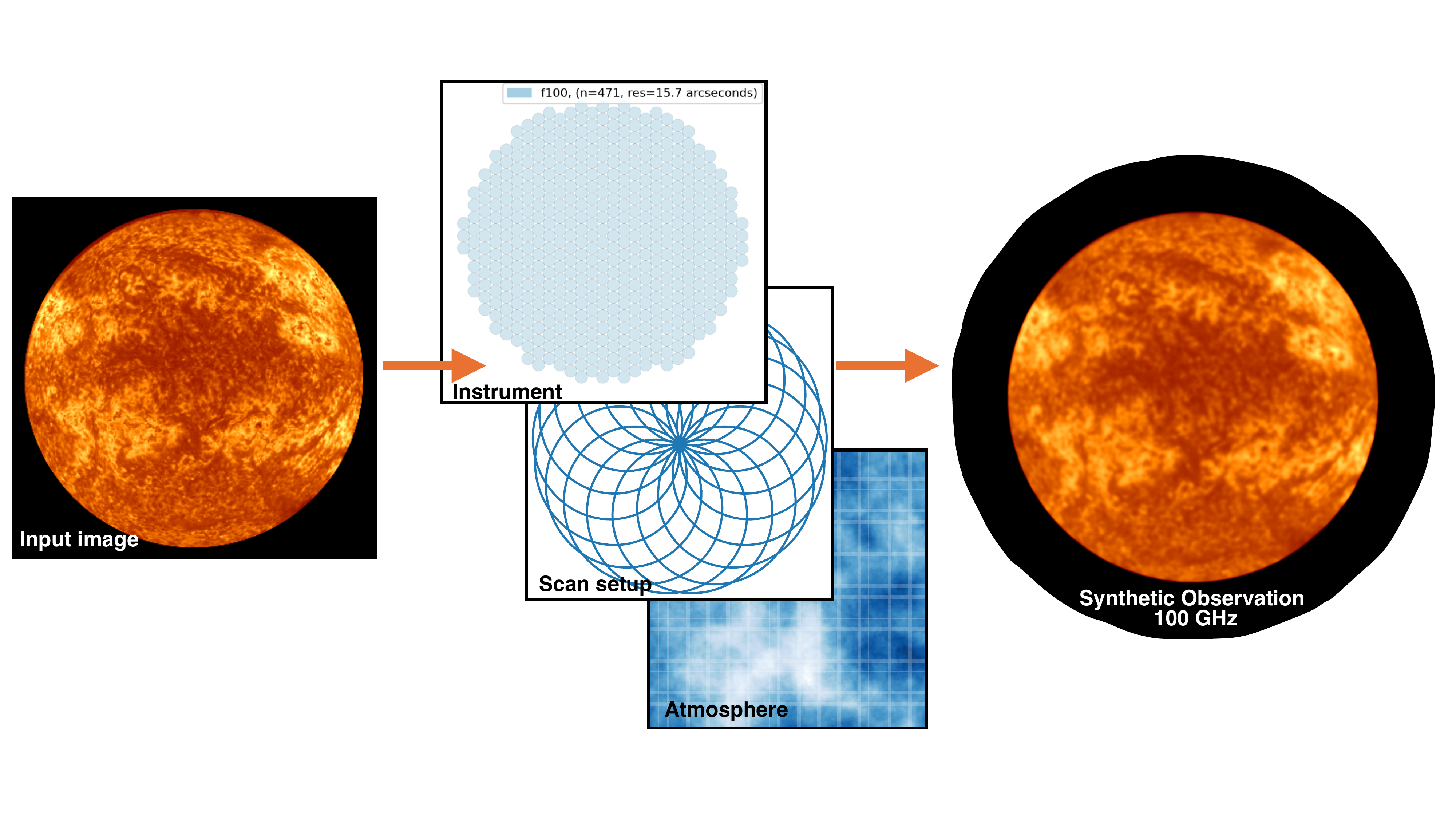}
    \vspace{-1.2cm}
    \caption{The various parts of the \texttt{maria} simulator. A synthetic observation is here created by simulating a scan of an input sky map, where both the instrument array, scanning strategy and atmosphere are highly customisable. The shown simulated scan is a 100 GHz scan utilising a relatively small \updated{circular} 471 detector instrument with a FOV of 0.1$^\circ$ at AtLAST, here scanning the disk in a double-circle pattern.}
    \label{fig:tmp_maria_method}
\end{figure*}

For investigating what instrumental setup would be required to carry out meaningful solar science with AtLAST in the coming decades, the \texttt{maria} tool is utilised to test instruments with a large range of properties, such as different number of detector elements, fields-of-view, but also different scanning strategies. An overview of the procedure of simulating AtLAST observations with \texttt{maria} is given in Fig.~\ref{fig:tmp_maria_method}.

\subsection{Realistic high-resolution input maps} \label{sec:input_maps}
To explore AtLAST's ability to observe the Sun with a range of different instrument setups and scan strategies, realistic synthetic observations have to be made, here using the \texttt{maria} simulator.
Realistic synthetic observations of the Sun at millimetre wavelengths require input in form of realistic maps of the Sun at these wavelengths at sufficiently high resolution. The latter should be better than the AtLAST beam across the entire considered wavelength range to successfully account for the impact of the AtLAST beam on the simulated observations.

Higher resolution millimetre observations of the Sun exist, for example from ALMA \citep{2016SSRv..200....1W, 2020A&A...635A..71W, 2022FrASS...9.7368B}, but the field of view of these observations is only of limited size, covering only a small fraction of the solar disk. To explore AtLAST's capability for full-disk mapping we need a full-disk input image, but no full-disk millimetre observations of the Sun exist with sufficiently high resolution. We therefore have to rely on observations at other wavelengths, mainly the ultraviolet, for input maps. 

However, full-disk maps in the ultraviolet are not perfect representations of the millimetre Sun, as slightly different heights in the atmosphere are probed. Nevertheless, they can act as sufficiently realistic reference maps, especially if one scales them to expected temperature levels and combine maps from different UV channels together to get a map better resembling the millimetre Sun. This was done by initially querying SDO AIA \citep{2012SoPh..275...17L, 2012SoPh..275....3P} maps from the 304 and 1600 Å channels at the same epoch of existing full-disk ALMA TP observations. Full-disk ALMA TP maps are available in ALMA Band 3, Band 6 and Band 7, in the frequency bands 84-116, 211-275 and 275-373 GHz respectively \citep{2019athb.rept.....R}. After convolving these UV maps to the ALMA primary beam in the respective frequency bands ($\sim 62.9\arcsec$, $\sim 25.9\arcsec$, and $\sim 19.4\arcsec$ at their respective central frequencies), and directly comparing them, it was found that the quiet Sun (QS) regions in ALMA maps best resembled those in AIA 304 Å, while the shapes of the active regions and plages (regions of enhanced emission in active regions) looked more similar to those observed in AIA 1600 Å.\updated{ We therefore construct the required millimetre full-disk maps by taking AIA 304 map, keeping the QS regions, but replacing the active regions with those in AIA 1600 Å multiplied with a scaling factor. This was done using a simple intensity mask in the 304 Å map, the threshold of which had to be adjusted slightly at different epochs to achieve the visually best reference map.}

Afterwards, the intensity distribution of the constructed map was scaled to follow the temperature distribution of the reference ALMA TP map, thus creating a reference input map that both resembles an actual millimetre observation of the Sun visually, but also in the distribution of the brightness temperatures. As the brightness temperatures on the Sun vary greatly across the frequency range of AtLAST, the input maps used at different frequencies were scaled separately to cover realistic temperature ranges. This was done by utilising the relation between wavelength and brightness temperature from \citet{2004A&A...419..747L} and scaling it to fit the mean brightness temperatures of available full-disk ALMA TP maps. 

\begin{figure*}
    \centering
    \includegraphics[width=1\linewidth]{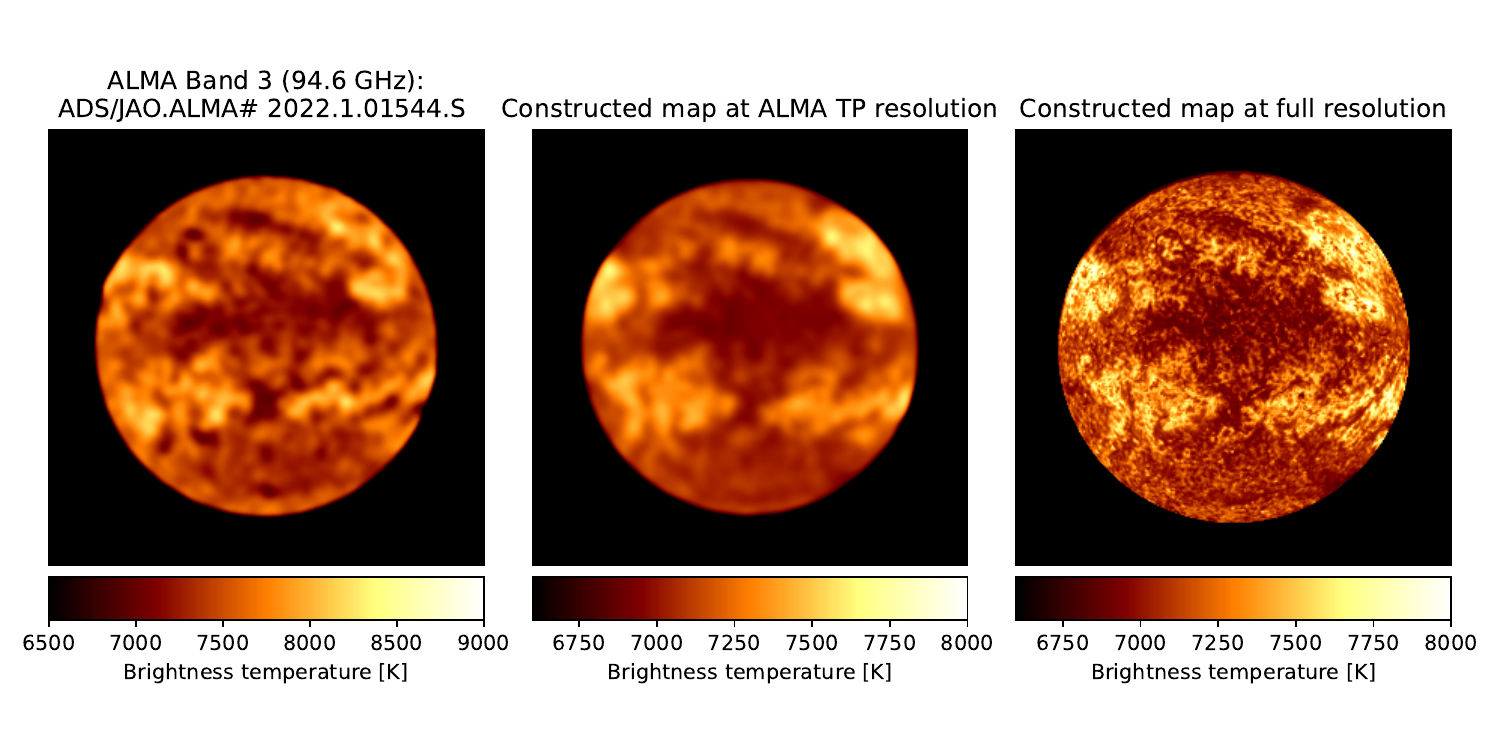}
    \vspace{-10mm}
    \caption{A comparison between an ALMA TP map in Band 3 taken on 11.03.2023 (ADS/JAO.ALMA\# 2022.1.01544.S) and a constructed map from SDO AIA data meant to resemble it. The constructed map is shown convolved to the ALMA TP beam for ease of comparison, and at its full resolution for further use in \texttt{maria}. It should be noted that the brightness temperature ranges are \updated{similar}, but not identical.}
    \label{fig:conv_comp}
\end{figure*}

An example of such a constructed map for the ALMA Band 3 taken on 11.03.2023 (ADS/JAO.ALMA\#2022.1.01544.S) is shown in Fig.~\ref{fig:conv_comp}, where the full-resolution map is shown, as well as one convolved to the beam of an ALMA TP antenna in band 3. This map is used as the base of all \texttt{maria}-simulations in this paper. Although only based on a single Band 3 ALMA TP map, the constructed map is used across all considered frequencies for AtLAST in Sections.~\ref{sec:results} and \ref{sec:discussion}. \updated{This map was chosen due to the presence of large active regions with complicated structures covering a broad range of spatial scales, making it an ideal test case for the scan's capability to reconstruct such complicated structures. Additionally, the parts of the map outside active regions are representative of quiescent regions, i.e. ``Quiet Sun'', therefore also testing the scan's capability to capture faint structures and thus also  maps at epochs closer to solar minimum.}

%This is sufficiently realistic for exploring the viability of using AtLAST for high-cadence full-disk scanning. 

\subsection{\texttt{maria}-simulated solar observations}

The process of creating synthetic observations in \texttt{maria} is thoroughly explained in \citet{2024OJAp....7E.118V}. The \texttt{maria} code works by defining an instrument, a telescope and a scanning pattern, each highly customisable with several parameters. The scan itself then creates time ordered data (TODs) for each detector, yielding the pointing coordinates (right ascension (RA) and declination (DEC), the brightness temperature of the scan, and the associated time stamp \citep{2024OJAp....7E.118V}. These TODs are then used to construct a synthetic map. 

The details of any future instrument at AtLAST are still undecided, but \citet{AtLAST_memo_4} has gathered estimates for the ranges of instrumental parameters for different types of instruments at AtLAST. As multi-chroic cameras were found to be a promising choice of instrument in \citet{2024ORE.....4..140W}, and \texttt{maria} is especially suited for simulated focal plane arrays, we here choose to focus our efforts on studying the range of instrumental parameters currently estimated for multi-chroic cameras at AtLAST. The range of parameters for multi-chroic cameras estimated by \citet{AtLAST_memo_4} for the AtLAST concept design \citep{2024SPIE13094E..28G, 2024SPIE13094E..4SP, 2024arXiv240218645M} are summarised in Table.~\ref{tab:range_tab}. 

As discussed in \citet{2024ORE.....4..140W}, a solar multi-chroic camera at AtLAST should at least have a frequency range from 90 GHz to 660 GHz, where it was argued that although going to frequencies up to $\sim$ 1 THz would provide higher angular resolution, adequate observing conditions might not occur as often, making these observations unreliable if one opts to observe the Sun periodically (perhaps daily) over a long time. The (theoretical) diffraction-limited angular resolution (in radians) of a single-dish telescope is a function of the observing frequency (or wavelength) and the aperture diameter, 
    $\theta = 1.22 \lambda / D = 1.22 c / (D\nu)$,
where $\lambda$ is the wavelength, $\nu$ is the frequency, and $D$ is the aperture diameter. Therefore, observing at 950 GHz instead of 660\,GHz with AtLAST, which would be diffraction limited, would result in an angular resolution of 1.59$\arcsec$, instead of the 2.29$\arcsec$ at 660\,GHz. For completeness we consider frequencies across AtLAST's frequency range of 30 - 950~GHz, as simulating high frequency operations at relatively poor seeing conditions might give some early insight into AtLAST's possible performance at these frequencies. 

\subsection{Considered parameter space}
One of the most crucial properties of a focal-plane array instrument regarding scanning extended astronomical bodies is its instantaneous FOV. This property is discussed in more detail in Sect.~\ref{sec:fov}, and depends on three parameters; the number of detectors, the spacing between the detectors, and the observing frequency. Observing at higher frequencies makes the beams of each detector smaller, meaning that a more dense configuration is needed to sufficiently sample a region on the sky. Therefore it is the highest frequencies that limit our multi-chroic instrument's observational capability, as one can reach the same instantaneous FOV much easier at lower frequencies with fewer detectors. 
%The synthetic observations presented in Sect.~\ref{sec:results} are therefore at AtLAST's highest frequency of 950~GHz, as if the instrument can successfully observe the solar disk at a certain cadence at 950~GHz, it could also achieve the same cadence at any lower frequency with even fewer detectors. 

For a first generation solar instrument at AtLAST we consider instruments with up to 50,000 detectors per frequency band, although more modest setups are also studied. For a second generation instrument, detector counts of up to 300,000 are considered \citep{AtLAST_memo_4}. These detectors could be spaced with detector spacings from 0.5 to 2.0 f$\lambda$, yielding different instantaneous FOVs. \updated{The f-number is here the ratio between the focal length and the aperture diameter, and $\lambda$ is the wavelength.}

\section{Results}
\label{sec:results}

\subsection{Instrumental properties}

\begin{table*}[ht!]
    \centering
    \begin{tabular}{ccccc}
        \hline
        \hline
        Instrumental property & Concept design range & Case A & Case B & Case C \\
        \hline
        Observing frequency [GHz]& ~30 - 950 & 100 - 950 & 100 - 670 & 100 - 950 \\
        \multirow{ 2}{*}{\updated{Pixel} count} & $\leq$ 50,000 per band (1st gen) & \multirow{ 2}{*}{1000} & \multirow{ 2}{*}{50,000} & \multirow{ 2}{*}{100,000} \\
        & $\leq$ 300,000 per band (2nd gen) & & & \\
        Pixel spacing [$f\lambda$] & 0.5 - 2.0 & 2.0 & 1.0 & 2.0 \\
        Field of view (diameter) [$^\circ$] & $\leq$ 2 & 0.03051 ($\sim 0.12$~R$_\odot$) & 0.146679 ($\sim 0.59$~R$_\odot$)& 0.30573 ($\sim 1.22$~R$_\odot$) \\
        \hline
    \end{tabular}
    \caption{Ranges of the considered instrumental properties, as well as the properties used in the three cases discussed in Sections \ref{sec:small_fov}, \ref{sec:int_fov}, and \ref{sec:large_fov}}
    \label{tab:range_tab}
\end{table*}
 
Finding the ideal properties of an instrument highly depends on the astronomical body of interest, the observing frequency, and also the required time cadence. For many sources of interest in the millimetre regime time cadence is of no concern, but when studying dynamic solar phenomena, having a sufficiently high cadence is crucial. To observe the Sun at millimetre wavelengths at sufficiently high cadence, an instrument with a very specific set of properties is required. In this section we introduce the properties we consider, followed by an analysis of how these 
%instrumental properties 
affect both the quality and time cadence of solar millimetre observations.  

\subsubsection{Field of view} \label{sec:fov}
When observing astronomical bodies with large apparent sizes such as the Sun in the millimetre regime, the FOV of the instrument is usually much smaller than the region of interest, thus introducing the need to scan the source. This impacts the cadence greatly, as scanning in a large and often complex pattern may take a long time. The scan time is dependent on the extent of the region one wishes to scan, the scan pattern, as well as the properties of the telescope itself such as its acceleration and velocity limits.
The most efficient scan pattern and the associated scan time is highly dependent on the instrument FOV, as one can permit leaving larger gaps between scan paths if the FOV is sufficiently large. Larger gaps between scan paths result in a shorter total scan path, and therefore a shorter scan time. From this it is evident that using large-FOV instruments can help push the scan times, and therefore also the cadence down to reasonable scales, suitable for observing dynamic solar phenomena. 

The FOV of an instrument is itself dependent on several other instrumental properties that would be useful to explore. For typical single-beam radio/millimetre antennas this FOV is just the primary beam diameter, which itself is dependent on the observing frequency and aperture size. However, with the advent of focal plane arrays, one can 
in practice populate the focal plane with large arrays of detectors filling significant portions of the field of view, and spanning multiple frequencies, with each detector's beam operating at our near the diffraction limit \citep{2024SPIE13094E..28G}. 
Such arrays have already been installed on facilities such as the Green Bank Telescope (GBT), namely the 223 element focal plane array MUSTANG-2 \citep{2014JLTP..176..808D}. The number of detector elements, henceforth referred to as pixels, and the spacing in-between them is then what sets the effective FOV of the instrument. A first generation multi-chroic instrument on AtLAST is expected to have up to 50,000 pixels per frequency band, while a second generation instrument is expected to host up to 300,000 pixels per band \citep{AtLAST_memo_4}.
These estimates of the number of detector elements act as upper limits to the theoretical range of pixel counts we consider in this study as shown in Table.~\ref{tab:range_tab}. As for the spacing between each pixel, that is as of now not clearly defined. As such, we operate with pixel spacings in the interval from 0.5 to 2.0 f$\lambda$, covering Nyquist sampling at 0.5 f$\lambda$ and the case where a full beam can fit in-between the beams of the pixels in the detector at a 2.0 f$\lambda$ spacing. This range of pixels spacings is shown in Table.~\ref{tab:range_tab}. The instrumental properties in question also impact the choice of suitable detector technology. A detector spacing of 0.5 f$\lambda$ is too short for typical feed horn-coupled arrays, but suitable for absorber coupled arrays. With spacings larger than 1 f$\lambda$, feed horn-coupled arrays are likely the most relevant \citep{2002ApOpt..41.6543G}. 

\subsubsection{Scan pattern} \label{sec:scan_patterns}
There are many different scan strategies implemented at millimetre facilities across the globe. At ALMA, full-disk maps of the Sun are created by scanning with a singular 12-m antenna in a double-circle pattern \citep{2017SoPh..292...88W}. As described in \citet{2017SoPh..292...88W} the double-circle pattern consists of minor circles with diameters of half the diameter of the scanned region, whose centres steadily move along a major circle of the same diameter but centred on the scanned region. An example of such a scan is depicted in Fig.~\ref{fig:tmp_maria_method}. \citet{2017SoPh..292...88W} mention that one of the strengths of such a pattern is that it allows the instrument to have a relatively steady velocity, so the acceleration is kept to a minimum. They mention that other methods, such as a Lissajous, requires sharp turns, and therefore higher acceleration and jerk. 

Lissajous scans are similar to double-circle scans in that they are also combinations of two sinusoidal functions, but of different amplitudes, periods and phases. \updated{The Lissajous daisy pattern is a special kind of Lissajous, where the instrument boresight scans the sky in a petal-like path, thus resembling a daisy. See \citet{2024OJAp....7E.118V} for an example using a Lissajous daisy scan to mimic MUSTANG-2 \citep{2014JLTP..176..808D} observations with maria.} 
%(see Fig.~\ref{fig:scan_patterns} for an example of a typical Lissajous daisy with a centre offset). 
This scan pattern, which is utilised at the GBT \citep{daisy_gbt, 2011ApJ...734...10K}, has radially increasing noise levels, usually having a well-observed central region and increasingly noise-dominated outer regions due to the decreased coverage \citep{2011ApJ...734...10K}. \updated{Similar to the double-circle pattern, the Lissajous daisy has the advantage of allowing the telescope to reach higher scan speeds than typical Lissajous patterns, without very sharp turns that require drastic accelerations \citep{2011ApJ...734...10K}.}

However, the efficiency of both the double-circle and the Lissajous daisy scan patterns are wholly dependent on the size of the region of interest, as well as the FOV of the instrument in use, and especially for large-FOV instruments it is plausible that more efficient scan patterns exist, especially for relatively small sky patches. This is believed to become relevant for the Sun, with its $\sim 0.5~^\circ$ apparent diameter as soon as the instrument FOV diameter approaches the solar radius. Using such a large instrument in a double-circle becomes redundant, as one ideally here could cover the whole solar disk by simply scanning in a circular path of a radius exactly half that of the solar radius, thus fully sampling the full disk in one simple path that could be scanned on very short time scales, thus resulting in very high cadence observations. This possibility is further explored in Sect.~\ref{sec:large_fov}. 

\subsection{Synthetic observations with different instrumental setups}
The following sections summarise the process of simulating synthetic scans of the solar disk. Three different instrumental setups with different scan strategies are discussed in detail. 
These setups are selected to cover the range of different possible properties for a 1st and a 2nd generation instrument with correspondingly small to large instantaneous FOV. The setups are named Case A - C, where the small FOV setup (Case A) is discussed in Sect.~\ref{sec:small_fov}, the intermediately sized FOV setup (Case B) in Sect.~\ref{sec:int_fov}, and the large FOV setup (Case C) is discussed in Sect.~\ref{sec:large_fov}. \updated{These three instrumental setups all have circular detector arrays, and are summarised in Table.~\ref{tab:range_tab}.} \updated{For setups with multiple frequency bands, separate detector arrays are setup for each of the four bands, equivalent to having separate arrays in the same focal plane or splitting the light into the considered frequencies before it reaches the individual detectors.}

\begin{figure*}[]
    \centering
    \begin{subfigure}[b]{0.32\textwidth}
       \centering
        \includegraphics[width=\textwidth]{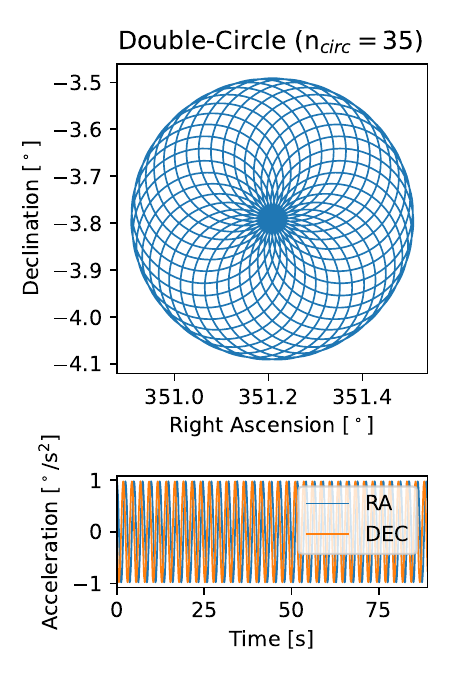}
        \label{fig:subfig_scan_A}
    \end{subfigure}
    \hfill
    \begin{subfigure}[b]{0.32\textwidth}
        \centering
        \includegraphics[width=\textwidth]{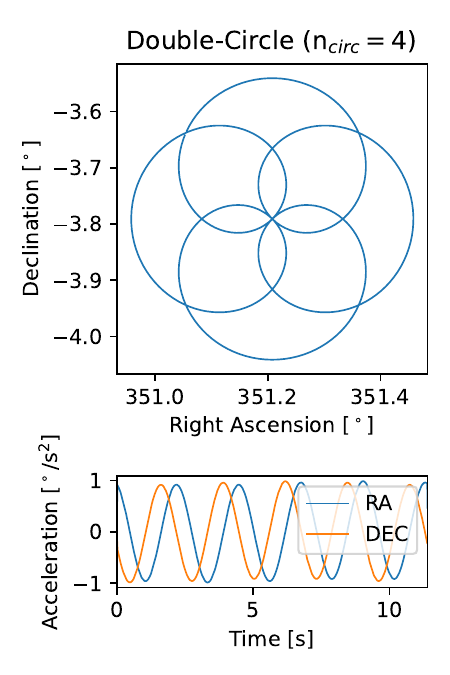}
        \label{fig:subfig_scan_B}
    \end{subfigure}
    \hfill
    \begin{subfigure}[b]{0.33\textwidth}
       \centering
        \includegraphics[width=\textwidth]{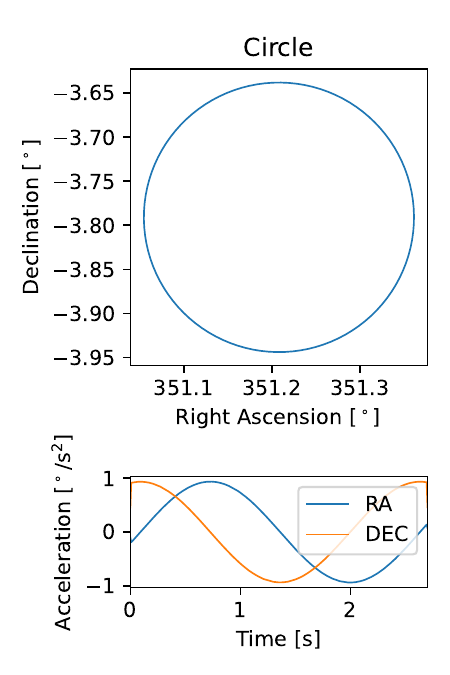}
        \label{fig:subfig_scan_C}
    \end{subfigure}

    \caption{a): Scan pattern and corresponding boresight acceleration for the small-FOV instrument discussed in Sect.~\ref{sec:small_fov}. b): Scan pattern and corresponding boresight acceleration for the intermediate-FOV instrument discussed in Sect.~\ref{sec:int_fov}. c): Scan pattern and corresponding boresight acceleration for the large-FOV instrument discussed in Sect.~\ref{sec:large_fov}.}
    \label{fig:scans_comp_fovs}
\end{figure*}

\begin{figure*}[hp!]
    \centering
    \includegraphics[width=1\linewidth]{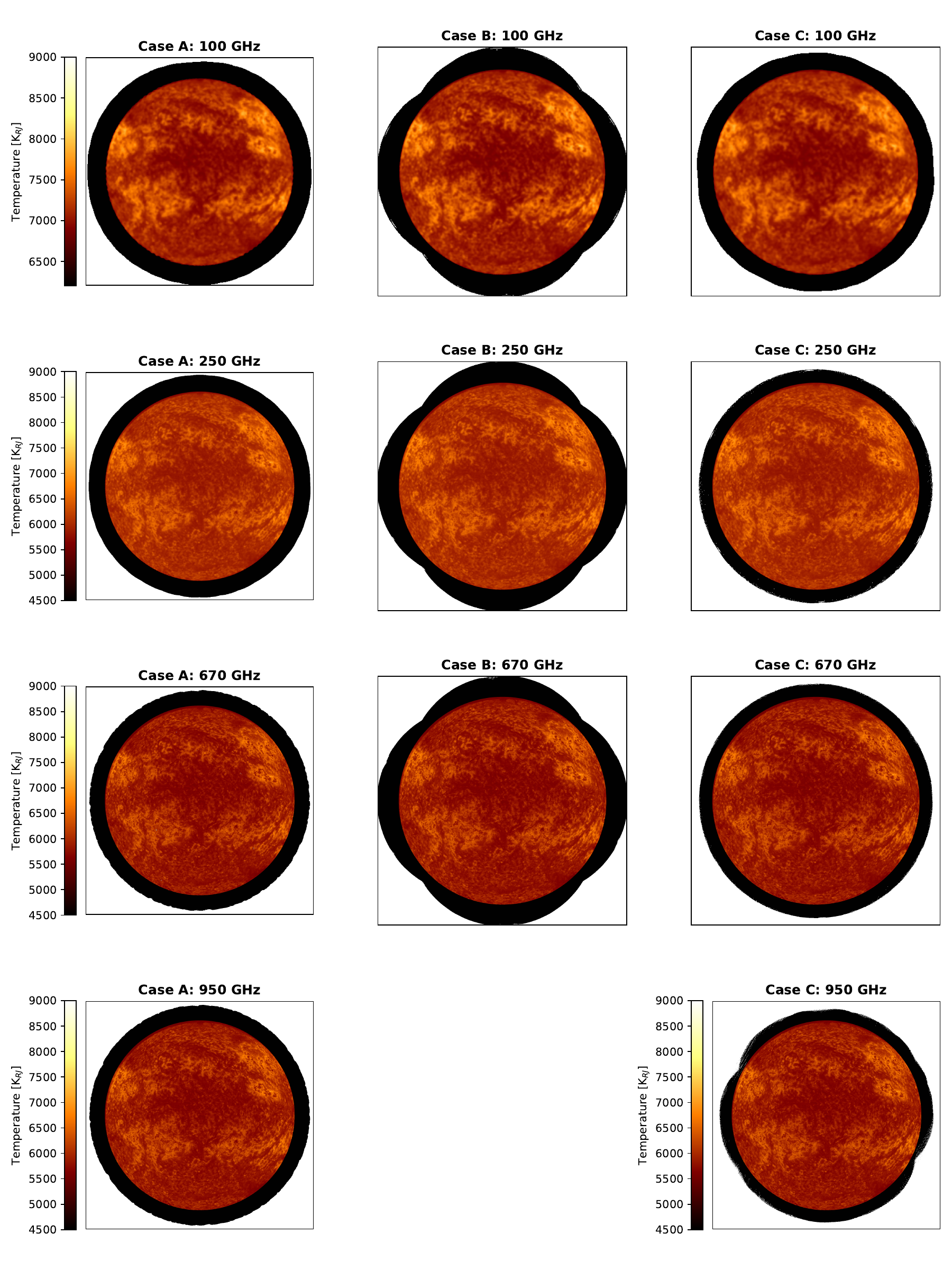}
    \caption{The synthetic observations resulting from the three cases discussed in Sect.~\ref{sec:small_fov}, \ref{sec:int_fov}, and \ref{sec:large_fov}. The simulations are without an atmosphere.}
    \label{fig:maps_comp_fovs}
\end{figure*}

\subsubsection{Scanning strategies for small FOV instruments (Case A)} \label{sec:small_fov}
Instruments with small fields of view (FOV) require dense and often complex scan strategies to sufficiently sample extended astronomical bodies such as the Sun. One such scan pattern is the previously mentioned double-circle as utilised at ALMA \citep{2017SoPh..292...88W}. The density of the double-circle pattern, or the number of secondary circles, depends on the FOV of the instrument in use. The larger the instantaneous FOV, the further apart each secondary circle can be, and the shorter the total scan path. It therefore follows that larger FOV instruments require less time to complete their scan path, and can therefore achieve higher cadence than instruments with smaller FOVs.

A small-FOV, circular instrument is here set up, with a conservative 1000 detectors per band, separated on a 2 f$\lambda$ spacing. This results in an instantaneous FOV diameter of $0.03051^\circ$ at 950~GHz, which translates to $\approx 0.12$ R$_\odot$. Four bands are chosen to cover most of AtLAST's frequency range, centred on 100, 250, 670, and 950~GHz. The highest frequency is here the most important, as the beams are the smallest. Therefore, only at 950~GHz are the 1000 detectors used, as at the lower frequencies less detectors are necessary to fill the same FOV. A \texttt{maria}-simulation is performed to uncover the constructed instrument's scanning capabilities. The parameters of the simulated instrument are shown in Table.~\ref{tab:range_tab} (Case A). 

To completely sample the full disk with such an instrument a tight double-circle is required, i.e. a high number of secondary circles have to be completed for each rotation around the primary circle. It was found that 36 minor circles were sufficient for sampling the whole solar disk with such an instrument. The scan pattern and the associated boresight acceleration can be found in Fig.~\ref{fig:scans_comp_fovs}a, where it is seen that the acceleration rapidly oscillates periodically, as expected. These acceleration components never pass AtLAST's acceleration limit of 1 $^\circ$/s$^2$ \citep{2024arXiv240218645M}. It was found that the required time needed to complete such a scan was $~89$~seconds. As discussed in Sect.~\ref{sec:scan_patterns}, larger FOV instruments would require fewer secondary circles per primary rotation for successful sampling and therefore less time to complete a scan. This is further explored in Sect.~\ref{sec:res_cadence}. The completed noise-less \texttt{maria}-simulated scans of the Sun in all four bands with this instrument and scan pattern are shown in the first column of Fig.~\ref{fig:maps_comp_fovs}, where it is found that the structures observed in the input map seem to be recovered at all frequencies. Note that the same input map was used for all four frequency bands, but that the brightness temperatures have been scaled to those expected at the considered frequencies. Therefore, the brightness temperatures are lower at higher frequencies, while the structures look the same (See Sect.~\ref{sec:input_maps}). Note that due to large beam sizes in the 100 and 250 GHz bands, very few detectors are required to fill the instrument FOV. With a 2 f$\lambda$ spacing, these frequencies are prone to leave gaps in the simulated maps, which requires post-processing in the form of smoothing to fill. These are done with a Gaussian of standard deviation of 1 pixel for each of these two cases, which result in a slightly worse effective resolution. In actuality, one could easily fill these gaps at low frequencies by having a more dense detector array only at low frequency, as one could still fill the FOV with reasonably few pixels. This would have to be studied further in a dedicated instrumental study.

\subsubsection{Scanning strategies for intermediate-FOV instruments (Case B)} \label{sec:int_fov}
A focal plane array of only 1000 pixels per frequency band is far below the current estimates for even a first generation instrument at AtLAST, and it is therefore interesting to test a more representative instrument. According to \citet{AtLAST_memo_4}, a first generation multi-chroic camera at AtLAST is expected to be able to host around 50,000 detector elements per frequency band. The intermediate-FOV instrument is therefore chosen to have 50,000 pixels, and to have them spaced closer together, now on a detector spacing of 1 f$\lambda$. Packing the detectors closer together results in a smaller instantaneous FOV, and to counteract this a maximum observing frequency of 670 GHz is here chosen. We then only consider the three bands centred on 100, 250, and 670~GHz. The parameters of the simulated instrument are shown in Table.~\ref{tab:range_tab} (Case B)%For a discussion of the feasibility of going to higher frequencies for solar observations, see Sect.~\ref{}. 

The  50,000 pixels observing up to $670$ GHz at 1 $f\lambda$ spacing resulted in an instrument with a field of view of $~0.146679$ degrees, almost $60\%$ of a solar radius. 
As before, we consider a double-circle scanning strategy, however, with a relatively large FOV, few secondary circles are required for sufficiently sampling the solar disk. It was here found that having only four minor circles per rotation around the major circle was sufficient for fully sampling the whole disk. The scan pattern is here shown in Fig.~\ref{fig:scans_comp_fovs}b with its associated boresight accelerations. With such a simple scan pattern, no drastic accelerations of the telescope are required, meaning that the full scan can be completed on short time scales. It was here found that the shortest time AtLAST could complete such a scan is $\sim 11.4$~seconds, almost seven times as fast as the small-FOV instrument in a much tighter double-circle pattern. This shows the 
large improvement offered by building instrumentation that fills more of the focal plane.
The resulting synthetic observations from this scan can be found in the second row of Fig.~\ref{fig:maps_comp_fovs}, where as before all major structures from the input image have been recovered.

\subsubsection{Scanning strategies for large FOV instruments (Case C)}\label{sec:large_fov}

The double-circle pattern has proved to be an efficient method of scanning the solar disk when utilising instruments with small to intermediately sized fields of view. However, as one gets to larger FOV instruments such a method would prove less efficient. With very large FOV instruments, one can get away with simpler scan pattern.

Assuming the instrument FOV diameter to be  greater than 1 R$_\odot$, simply scanning in a circular motion would be sufficient to observe the full solar disk. Such an instrument would then be utilised in circular scans centred on the solar disk with a radius of half the instrument FOV, and thusly cover the full disk on possibly very short time scales. 
Such a case is considered, and we here consider the case where one wishes for the entire solar disk and $0.2 R_\odot$ off-limb to be fully sampled, meaning that the area we want to scan has a diameter of 2.4 R$\odot$. Using an instrument with a FOV of 1.2 R$_\odot$ and scanning along a circle of radius 0.6 R$_\odot$ will then cover the full disk, assuming the sample rate is sufficiently high and that the instrument FOV is sufficiently filled with detector elements to not leave gaps. An advantage of this scanning method is its simplicity, as it requires very little stress on the telescope, as the acceleration is kept low and slowly varying. 

A \texttt{maria}-simulation of such a scan is here created. As in the small-FOV case we set the highest observing frequency to be 950~GHz, and consider the same bands as in Sect.~\ref{sec:small_fov}. To keep the number of required detector elements reasonable we set the  pixel spacing to be 2~f$\lambda$. We allow the instrument to have 100,000 pixels per band, twice that of the first generation estimate, but only a third of the second generation estimate. It is therefore not wholly unlikely that such an instrument could be built by the time AtLAST is observation-ready. This instrument then has a FOV of 0.30573$^\circ$ ($\sim$1.22 R$_\odot$). The parameters of the simulated instrument are shown in Table.~\ref{tab:range_tab} (Case C). This instrument is then set up to scan along a circular path of radius 0.61~R$_\odot$, which it completes in about 2.7 seconds. The scanning path and the associated boresight acceleration can be found in Fig.~\ref{fig:scans_comp_fovs}c. 

The synthetic AtLAST observations resulting from this scan are shown in the third row of Fig.~\ref{fig:maps_comp_fovs}, where as for the two previous cases we find the structures in the input image to be recovered. As for case A in Sect.~\ref{sec:small_fov}, post-processing in the form of Gaussian smoothing with a standard deviation of 1 pixel was required to fill minor gaps in the map at 100~GHz. 

\newpage
\subsection{Cadence of observations} \label{sec:res_cadence}
Having seen that the scan time required to fully sample the solar disk is highly related to the FOV of the instrument in question, and therefore also the number of detector elements, their separation and the observing frequency, the relation between these could prove useful in deciding on the final instrumental properties of a future solar instrument for AtLAST. Having such a relation would allow one to directly find the required instrumental properties to achieve a desired time cadence at a desired upper frequency. 

\begin{figure*}[htp!]
    \centering
    \includegraphics[width=1.0\linewidth]{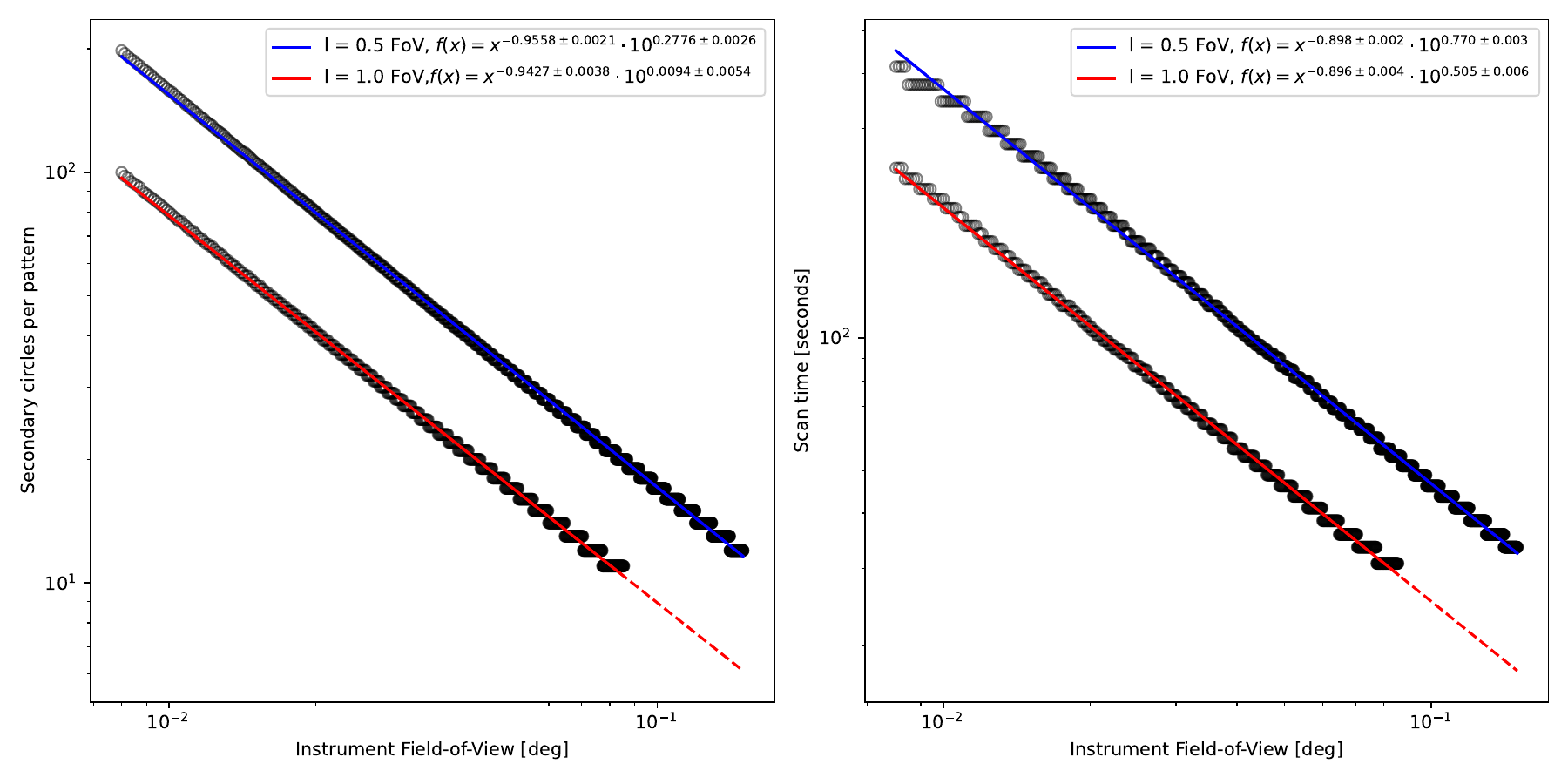}
    \caption{The number of secondary circles per pattern in the double-circle as a function of instrument field of view, in addition to the total scan time, and therefore cadence, as a function of field of view for two different sampling lengths, 0.5 and 1.0 the instrument FOV. A blue solid line shows the fitted function to the l = 0.5 FOV data, with a blue dashed line showing the extension of this fit outside the data range. Similarly, a red solid lines shows the function fitted to the l = 1.0 FOV data, with a red dashed line showing the extension of this function.}
    \label{fig:circle_time}
\end{figure*}

To arrive at such a relation, a double-circle scanning pattern was initially assumed, as the number of minor circles per pattern could directly be related to both the needed scan time to complete such a pattern, but also instrument FOV. Estimating the number of minor circles required to fully sample a region of a given size with a specified instrument could therefore yield a relation between the total needed scan time and the instrument FOV. To cover each point in a sky patch at least once, it is sufficient that the minor circles in the double-circle pattern are spaced out in such a way that gaps in-between them correspond to the diameter of the instrument FOV. The full region in-between two minor circles would thusly be sampled fully by two consecutive minor circle paths. The gap in-between two consecutive minor circles is called the sampling length \citep{2017SoPh..292...88W}, and in the case described here is equal to the diameter of the instrument's FOV, $l = 1.0$ FOV. Lowering the sampling length to $l = 0.5$ FOV \updated{means that} regions in-between two consecutive minor circles \updated{are covered twice, ensuring better sampling. This could be especially useful if the detector array is sparsely filled with pixels (detector elements)}.

To find a relation between the instrument's instantaneous FOV and the total scan time, the size of the sky patch was set to be a circular region of diameter 2400~$\arcsec$, the same size as the region scanned by ALMA TP antennas when mapping the solar disk \citep{2017SoPh..292...88W}. It was chosen to cover the full disk and a region of over 200$\arcsec$ off-limb, advantageous both for calibration and for observing solar prominences \citep{2016ApJ...833..141G, 2018ApJ...853...21G, 2017SoPh..292...88W}. A wide range of instrument FOVs were then set-up, and the required number of minor circles per pattern computed, both with sampling lengths of one and a half the diameter of the instrument FOV. This resulted in the relations:
\begin{align}
    N_{\rm circles, l=0.5} = \rm{FOV}_{\rm inst}^{-0.951 \pm 0.003} \cdot 10^{0.285 \pm 0.004},\,\, l = 0.5\,\,\rm{FOV}_{\rm inst},\\
    N_{\rm circles, l=1.0} = \rm{FOV}_{\rm inst}^{-0.943 \pm 0.006} \cdot 10^{0.008 \pm 0.008},\,\, l = 1.0\,\,\rm{FOV}_{\rm inst},
\end{align}
where FOV$_{\rm inst}$ is the FOV of the instrument, $N_{\rm circles, l=0.5}$ is the number of required minor circles with a sampling length of $l = 0.5$ FOV$_{\rm inst}$, and $N_{\rm circles, l=1.0}$ is the number of required minor circles with a sampling length of $l = 1.0$ FOV$_{\rm inst}$.
These relations are shown in Fig.~\ref{fig:circle_time}. 

Knowing the acceleration and velocity limits of AtLAST, we can directly relate the number of required secondary circles to the needed scan time, and therefore the observational cadence. With a acceleration limit of 1~$^\circ$/s$^2$ and a velocity limit of 3~$^\circ$/s \citep[see Table.1 in][for an overview of AtLAST's technical requirements]{2024arXiv240218645M}, the following relations were found:
\begin{align}
    t_{\rm scan, l=0.5} = \rm{FOV}_{\rm inst}^{-0.893 \pm 0.004} \cdot 10^{0.778 \pm 0.004},\,\, l = 0.5\,\,\rm{FOV}_{\rm inst},\\
    t_{\rm scan, l=1.0} = \rm{FOV}_{\rm inst}^{-0.897 \pm 0.006} \cdot 10^{0.504 \pm 0.009},\,\, l = 1.0\,\,\rm{FOV}_{\rm inst},
\end{align}
where $t_{\rm scan, l=0.5}$ and $t_{\rm scan, l=1.0}$ denote the needed scan times at sampling lengths of $l = 0.5$ $\rm{FOV}_{\rm inst}$ and $l = 1.0$ $\rm{FOV}_{\rm inst}$ respectively. These relations are shown in the right panel of Fig.~\ref{fig:circle_time}. As expected, the instrument FOV has a 
large impact on the scan time, and it can be seen that even going to small to intermediately sized FOVs would greatly reduce the required scan time. 

As discussed in Sect.~\ref{sec:fov}, the instrument's instantaneous FOV is itself dependent on several instrumental properties, here the observing frequency, the number of pixels and the pixel spacing. It is therefore possible to instead relate the scan time to the number of pixels at set observing frequencies and with specified pixel spacings. However, note that the assumption of the full instantaneous FOV being sufficiently filled with detectors to instantaneously cover the sky region it passes over still applies. This only applies to setups where we have sufficient numbers of detector elements to not leave gaps after the full detector array has passed the sky patch. We therefore do not expect any found trends to be applicable at large detector spacings when the number of detectors is low. However, we argue that the relations are still valid at detector counts currently considered for AtLAST. 
As a first generation multi-chroic camera at AtLAST is expected to have four frequency channels, we do this for four different frequencies, and for the three pixel spacings 0.5, 1.0, and 2.0~f$\lambda$ for sampling lengths of $l = 0.5$ and $1.0$ FOV. The four frequency channels were chosen to be centred on 100, 250, 670, and 950~GHz. The three first bands are then counterparts to Band 3, 6, and 9 at ALMA \footnote{\href{https://www.eso.org/public/teles-instr/alma/receiver-bands/}{https://www.eso.org/public/teles-instr/alma/receiver-bands/}}, while the last is meant to check the feasibility of using AtLAST for high-cadence and high-frequency observations of the full disk. 

\begin{figure*}
    \centering
    \includegraphics[width=0.8\linewidth]{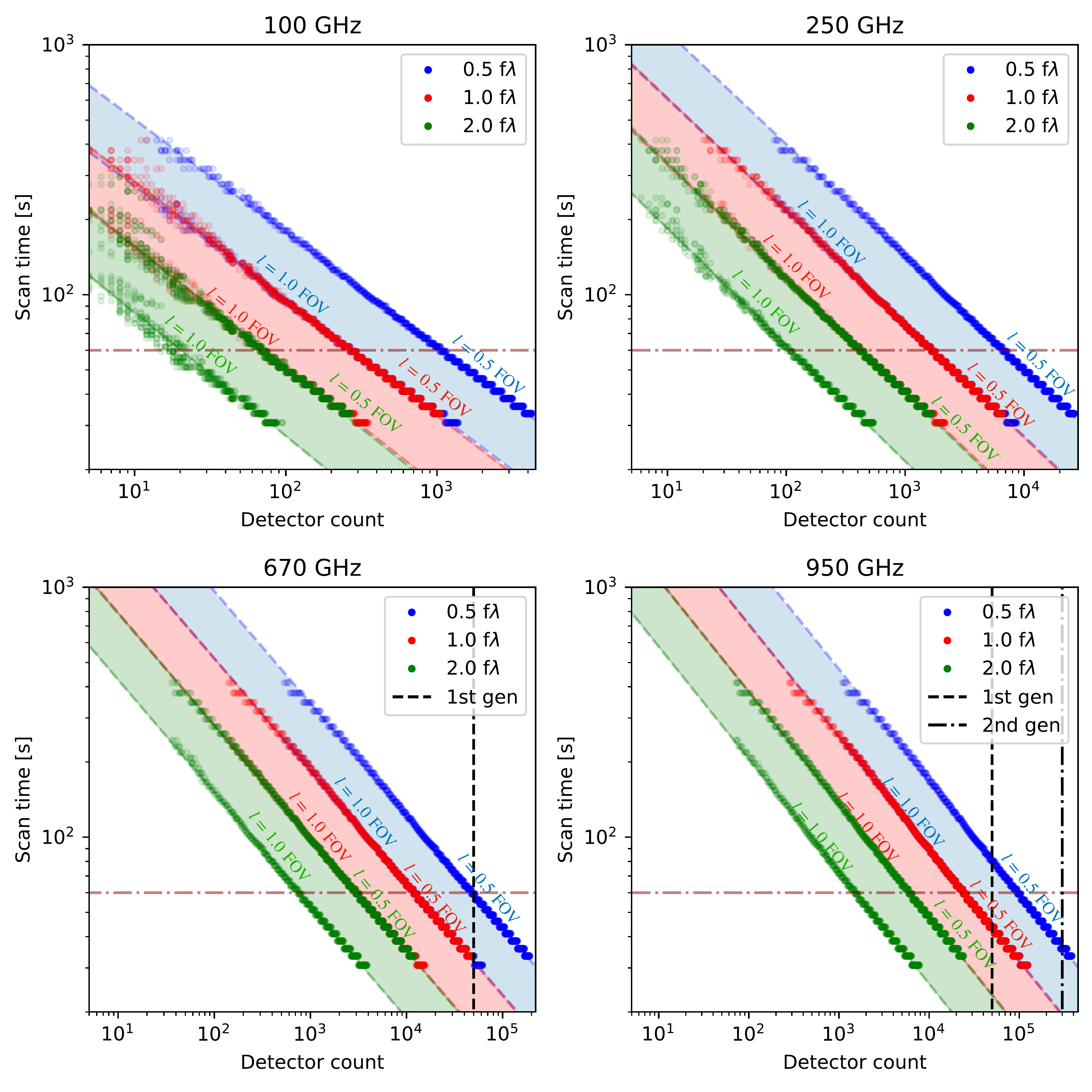}
    \caption{Time cadence as a function of the number of detector elements (pixels) with different pixel spacings for four custom frequency bands. Black dashed lines denote the upper limit on the number of detector elements per frequency band estimated for a first- and second generation multi-chroic camera at AtLAST. A red dashed line shows where the scan time equals 1 minute.}
    \label{fig:stime_npix_rel}
\end{figure*}

The relations between the scan time and the number of pixels are shown in Fig.~\ref{fig:stime_npix_rel}, where black dashed lines show the expected maximum number of pixels for a 1st and 2nd generation multi-chroic instrument \citep{AtLAST_memo_4}.
Blue, red, and green are here used to show different pixel spacings. One can here see the large effect multi-pixel instruments can have on the cadence of full-disk solar observations, as even going to the 50,000 pixel first generation instrument results in sub-minute cadence across all the four frequency bands, except in 670 and 950~GHz with a pixel spacing of 0.5~f$\lambda$. The improvement in cadence over ALMA TP full-disk maps is then on the order of ten, in addition to a four times higher angular resolution. \updated{It is also found that scanning at twice the sampling length (i.e., $l = 0.5$ instead of $1.0$ FOV) impacts the scan time the same way as keeping the sampling length the same and doubling the spacing in-between each pixel in the instrument (i.e., $1.0$ instead of $0.5$ f$\lambda$). Many of the data points are therefore superimposed on each other, making them difficult to distinguish. To make this easier, the area between first-order log-log fits to the two different sampling lengths has been shaded to the appropriate colour associated with the pixel spacing.}

\newpage
\section{Discussion} \label{sec:discussion}

\subsection{Required cadence} \label{sec:disc_cadence}

With the detection of solar flares with ALMA being both rare and notoriously difficult \citep{2021ApJ...922..113S, 2023A&A...669A.156S}, little is known about the time scales of 
flare emission at millimeter wavelengths. 
The interferometric ALMA observations of a flare by \citet{2021ApJ...922..113S} exhibited a peak that lasted $\sim$5~minutes. To sufficiently sample the evolution of such a flare with a solar instrument at AtLAST, observations with a (sub-)minute cadence would be required. The small-FOV instrument (case~A) considered in Sect.~\ref{sec:small_fov} with a cadence of $~89$~seconds ($\sim$ 1.48 minutes) would only be able to complete 3 observations during the flare. This would be sufficient to detect the flare, but insufficient for adequately studying the evolution of the flare. It is evident that for studying the time evolution of micro- or nano-flares, higher cadence is required. To achieve this, an instrument with a larger FOV is required, as the intermediate- and large FOV instruments described in Sect.~\ref{sec:int_fov} and \ref{sec:large_fov}, could achieve cadences of $\sim$11.4 and $\sim$2.7~seconds respectively. With such a high cadence, the peak of the microflare reported by \citet{2021ApJ...922..113S} would be sampled with $\sim$25 (case~B) and $\sim$111 time steps (case~C),  respectively.

The trends in scan time as a function of detector counts for the four frequency bands considered in Sect.~\ref{sec:res_cadence}, and shown in Fig.~\ref{fig:stime_npix_rel}, were made with the scan pattern spaced out in such a way as to ensure each patch of sky is passed at least twice by the instrument, i.e. a sampling length of half the FOV diameter of the utilised instrument ($l = 0.5$ FOV). Without more detailed  knowledge of the time scales needed to observe the shortest lived dynamic phenomena in the chromosphere, it is difficult to decide on a lower limit cadence and thus, via the relation between cadence and pixel count, the minimum number of pixels required for a set pixel spacing at each observing frequency. It should be noted that 
typical periods of quasi-periodic pulsations that occur during flares range from a fraction of a second to several minutes \citep[see, e.g.,][and references therein]{1969ApJ...155L.117P, 2016SSRv..200....1W}.

The number of pixels required for the considered frequency bands at different pixel spacings to reach a one minute cadence are given in Table.~\ref{tab:npix_min_cad}. We find that the number of required pixels is highly dependent on both the observing frequency and the spacing between pixels, as expected. 
A first generation multi-chroic camera with 50,000 pixels per frequency band, in accordance to the specifications as provided by the AtLAST instrumentation group \citep{AtLAST_memo_4}, could achieve a cadence of 1~min across the considered frequency bands and pixel spacings (see Table~\ref{tab:npix_min_cad}), except with a spacing of $0.5$~f$\lambda$ observing at 950~GHz. Interestingly, less than 25,000 pixels are required to observe up to 950~GHz with a spacing of 1~f$\lambda$, well within the estimates of a 1st generation instrument while still achieving a high cadence.

\updated{If one instead considers an instrument consisting of separate detector arrays of monochroic pixels for each frequency band, we find that all detector arrays could be at a $0.5$ f$\lambda$ spacing except for 950~GHz, where the pixels would have to be spaced further apart ($1.0$ f$\lambda$) for the required number of pixels to be within the estimate for a 1st generation instrument. }

%This is encouraging for an instrument that employs dedicated (monochroic) detectors for different frequency ranges. The detector for the highest covered frequency range requires the highest number of pixels, thus setting the limit for the overall instrument. On the other hand, the detectors for the lower frequencies would require fewer pixels while still able to reach a 1-min-cadence.  

\begin{table}[b!]
    \centering
    \begin{tabular}{ccccc}
        \hline 
        \hline 
        Pixel spacing & 100 GHz & 250 GHz & 675 GHz & 950 GHz \\
        \hline
        0.5 f$\lambda$ & 1095 & 6784 & 48,653 & 97,768 \\
        1.0 f$\lambda$ & 280 & 1708 & 12,165 & 24,464 \\
        2.0 f$\lambda$ & 74 & 433 & 3069 & 6108 \\
        \hline
    \end{tabular}
    \caption{The required number of pixels \updated{(detector elements)} for the considered frequency bands to achieve a one minute cadence. One can here see the large impact the pixel spacing has on the required pixels.}
    \label{tab:npix_min_cad}
\end{table}

\subsection{Efficient scan patterns} \label{sec:disc_scan_patterns}

We have seen that the double-circle scan pattern can be suitable for efficiently scanning sky patches, at least with instruments of reasonably small FOVs. As seen for Case C (see  Sect.~\ref{sec:large_fov}), when the FOV of the instrument becomes sufficiently large, scanning in an extended and often complex scan pattern is no longer as efficient. Instead, a simple scan pattern like a simple circular path is sufficient. The point where the FOV becomes sufficiently large as to warrant a change of scan strategy depends only on the size of the sky region of interest. As for the circular region of radius $2400\arcsec$ considered in Sect.~\ref{sec:res_cadence}, corresponding to the extent of the solar disk and a 200$\arcsec$ off-limb region as observed by ALMA \citep{2015ASPC..499..347P}, a circular instrument with an FOV diameter of 1200$\arcsec$ would be large enough to justify the use of a simple circular scan. The scan time for this FOV is independent of frequency, as long as the FOV is sufficiently filled with detectors at the respective frequencies. 

\updated{As seen in Table.~\ref{tab:single_circle_det_count}, assuming seperate monochroic detector arrays for each frequency, this results in much fewer pixels being required to fill the sufficiently large FOV to make the double-circle pattern inefficient at lower frequencies as compared to higher frequencies.} Table~\ref{tab:single_circle_det_count} shows the number of required detectors to fill the 1200$\arcsec$ diameter FOV which could be used in single-circle scans at different observing frequencies and detector spacings. It is here found that, if we want to push for observing at 950~GHz and utilise a single-circle scan strategy, detector spacings of less than $2.0$ f$\lambda$ would not be viable, as they would require detector counts above the current estimate for even a second generation instrument at AtLAST \citep[$\leq$ 300,000,][]{AtLAST_memo_4}.

\begin{table}[b!]
\begin{tabular}{l|llll}
\hline
\hline
Obs. Freq.       & \multirow{2}{*}{100 GHz} & \multirow{2}{*}{250 GHz} & \multirow{2}{*}{670 GHz} & \multirow{2}{*}{950 GHz} \\ \cline{1-1}
Pixel. Spacing &                          &                          &                          &                          \\ \hline
0.5 f$\lambda$   & 21 080                   & 131 645                  & 945 622                  & 1 901 240                \\
1.0 f$\lambda$   & $\,\,$ 5 279                    & $\,\,$ 32 943                   & 236 420                  & $\,\,\,$ 475 305                  \\
2.0 f$\lambda$   & $\,\,$ 1 320                    & $\,\,\,\,\,$ 8 223                    & $\,\,$ 59 108                   & $\,\,\,$ 118 853                  \\ \hline
\end{tabular}
\caption{Number of \updated{pixels} required to fill a focal plane array with a diameter equivalent to $\sim 0.33$ R$_\odot$ at different observing frequencies and detector spacings.}
\label{tab:single_circle_det_count}
\end{table}

\newpage
\subsection{Scan coverage}

\begin{figure}
    \centering
    \includegraphics[width=1\linewidth]{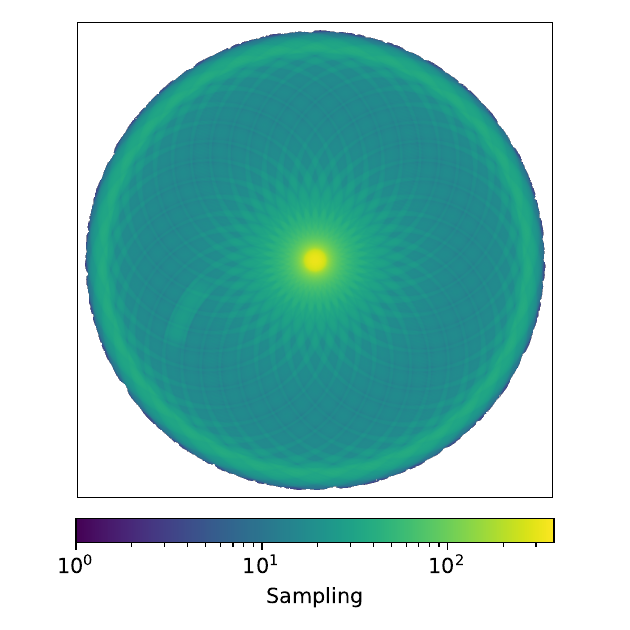}
    \caption{Sampling map of the 950~GHz synthetic observations using the Case A instrument discussed in Sect.~\ref{sec:small_fov}. \updated{This shows how many times each sky region is sampled during the scan.} The sampling rate is here 3750~Hz.}
    \label{fig:sampling_A_950}
\end{figure}

The considered scan patterns and strategies have already been shown to produce maps at significantly different cadences. However, the aspect of scan coverage has not been discussed in detail. From the double-circle scan path in Fig.~\ref{fig:scans_comp_fovs}a it is evident that this pattern is heavily weighted towards its centre. This is more clearly shown in the resulting weight map resulting from the 950~GHz scan as detailed in Sect.~\ref{sec:small_fov}, shown in Fig.~\ref{fig:sampling_A_950}. Note that the centre and the neighbouring regions are observed about many times as often as the bulk of the disk. However, it is noted that from immediately outside the central region the sampling is almost flat, and therefore uniform radially out to the limb. The double-circle pattern thus scans most of the solar disk with uniform coverage, with the downside of having a heavily weighted centre. However, this weighting towards the centre might prove useful when attempting to quantify (and correct for) the impact of Earth's atmosphere on the observations. \citet{2017SoPh..292...88W} states that an oversampled centre has the advantage that the fluctuations in the transmission of Earth's atmosphere can be tracked, and simply removed from the maps under the assumption that the centre's brightness temperature should remain stable on short time scales. Especially at higher frequencies this is believed to become important. 

A Lissajous daisy, as used by the Green Bank Telescope (GBT) \citep{daisy_gbt, 2011ApJ...734...10K}, has the advantage of a near-uniformly sampled central region when applying a centre-offset. However, the uniformly sampled region is often small compared to the size of the full scan pattern, meaning that the total scan path is long, requiring long scan times to achieve a uniformly sampled central region large enough to cover the solar disk. Nevertheless, the Lissajous daisy is an option that should be explored, in combination with other scan strategies. However, the suitability of the considered scan strategies for solar observing with AtLAST will depend on the details of the employed instrument. 

When considering \updated{instruments with large and filled FOVs}, it is expected that the scan pattern has less of an impact on both the image sampling and the resulting cadence, as one can fill the solar disk on short time scales simply by following a simple and often short scan path. However, for smaller instruments, as seen in Sect.~\ref{sec:small_fov}, a larger and more complex scan strategy is required to fully sample the solar disk, thus making the choice of a suitably efficient scan path all the more important. This both affects the sampling of the source, but also the resulting cadence. 

\begin{figure*}[hp!]
    \centering
    \includegraphics[width=.95\linewidth]{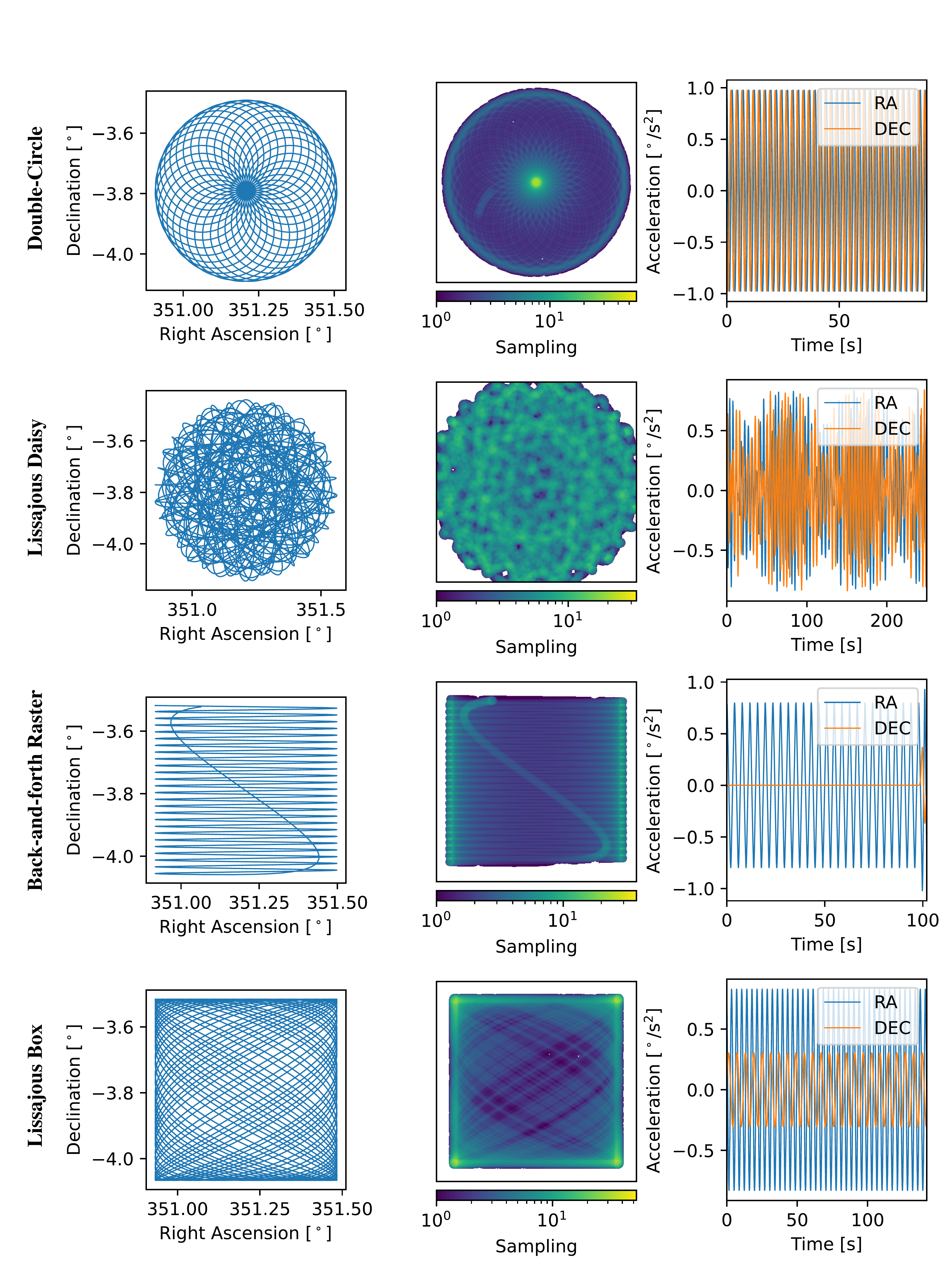}
    \caption{A comparison of four different scan strategies; double-circle, Lissajous daisy, back-and-forth raster, and a Lissajous box. The scan patterns are shown in the first column, their respective weight maps in the second column, and the boresight acceleration in the third column.}
    \label{fig:scan_comp_fig}
\end{figure*}

We here compare four scan strategies that could prove relevant for solar observing at AtLAST, namely the double circle, the Lissajous daisy, the raster, and the Lissajous box. For the comparison of these scan strategies, we consider a small instrument with  1000~pixels observing at 950~GHz. With a detector spacing of 2 f$\lambda$ this corresponds to a FOV of $\sim 0.03051^\circ$. The resulting tight scan patterns are shown in the left-most row in Fig.~\ref{fig:scan_comp_fig}. Note that the scan time for fully sampling the solar disk varies greatly between the strategies: 89\,s for the double-circle, 102\,s for the back-and-forth raster, 142\,s for the Lissajous box, and 250\,s for the Lissajous daisy pattern. The reason for the relatively long scan time needed for the Lissajous daisy to fully sample the whole solar disk is its tendency to miss small patches on the disk, so that it takes long before all patches have been covered. 
We conclude that the other three scan patterns provide a more even and thus efficient sampling. 
This is especially true \updated{for} the back-and-forth raster and the double-circle scanning the disk either from top to bottom or radially. These two patterns also have the advantage that in a possible malfunction that halts the scanning, the mid-scan data would still be usable, as they cover connected and fully sampled regions on the Sun. 

\begin{figure}
    \centering
    \includegraphics[width=1\linewidth]{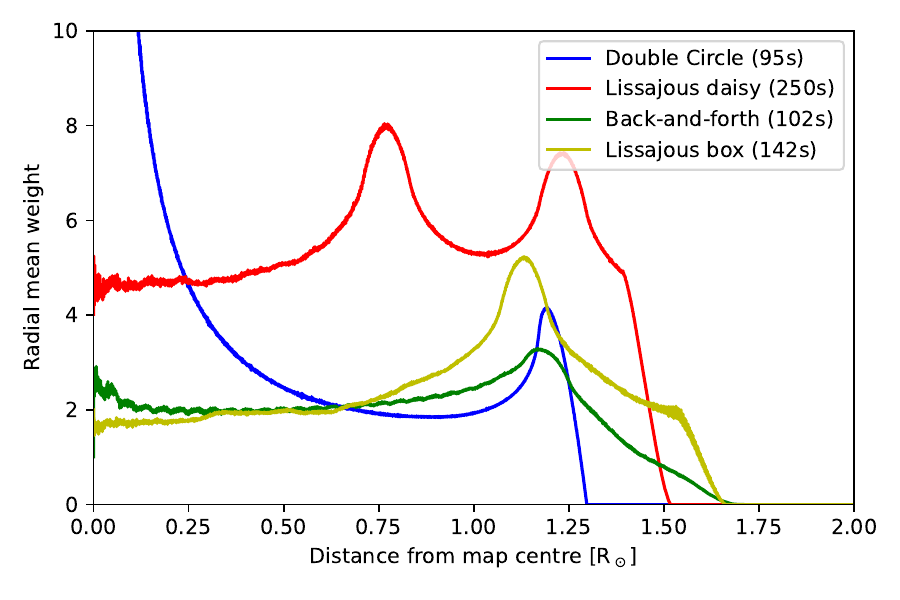}
    \caption{Radial mean sampling for the four considered scan strategies with a modest 1,000 detector instrument.}
    \label{fig:sampling_sfov}
\end{figure}

The weighting/sampling, being the number of times each pixel in the map are sampled by the detector, of the four scan strategies are shown in the second row of Fig.~\ref{fig:scan_comp_fig}. The radial mean weights of these maps are shown in Fig.~\ref{fig:sampling_sfov}. As expected, the double-circle is found to greatly oversample its centre compared to the other scan strategies, before reaching a near-uniform sampling of the bulk of the solar disk. Notably, both the back-and-forth raster and Lissajous box show near-uniform sampling across the full disk. The radial mean weight/sampling for the Lissajous daisy is significantly higher than for all other strategies. This is due to the previously mentioned issue of certain small patches remaining unsampled for extended periods. 
Consequently, the scan time becomes very long, and many regions on the Sun are sampled multiple times before all small patches are successfully sampled.

\begin{figure}
    \centering
    \includegraphics[width=1\linewidth]{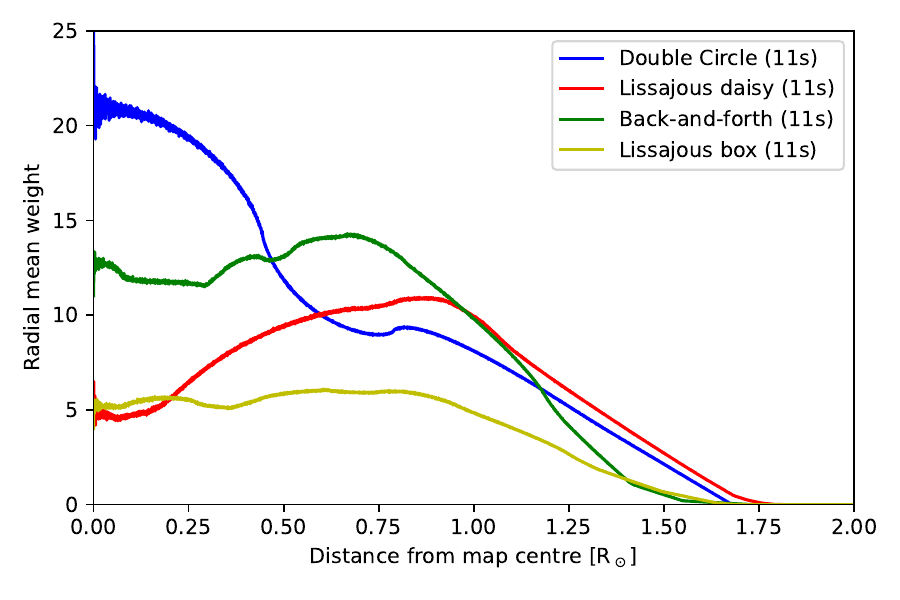}
    \caption{Radial mean sampling for the four considered scan strategies with a large 50,000 detector instrument.}
    \label{fig:sampling_lfov}
\end{figure}

\subsection{Quantifying the quality of synthetic observations}

In this work, we have created a number of synthetic AtLAST observations of the Sun with different instrumental setups, and thoroughly compared their associated scans and scan times to each other to give insight into AtLAST's future solar observing abilities. Even though this is sufficient to say something about AtLAST ideal high cadence, the quality of the individual scans should also be quantified, as high cadence observations are of little value if the scans are incapable of resolving the spatial scales present in the input map. To quantify the quality of the synthetic observations discussed in Sect.~\ref{sec:small_fov}, \ref{sec:int_fov}, and \ref{sec:large_fov}, we utilise spatial power spectra, as comparing these spectra to that of the input map will give insight into the individual scans' ability to recover the present spatial scales.  

\begin{figure*}[ht!]
    \centering
    \begin{subfigure}[b]{1\textwidth}
       \centering
        \includegraphics[width=\textwidth]{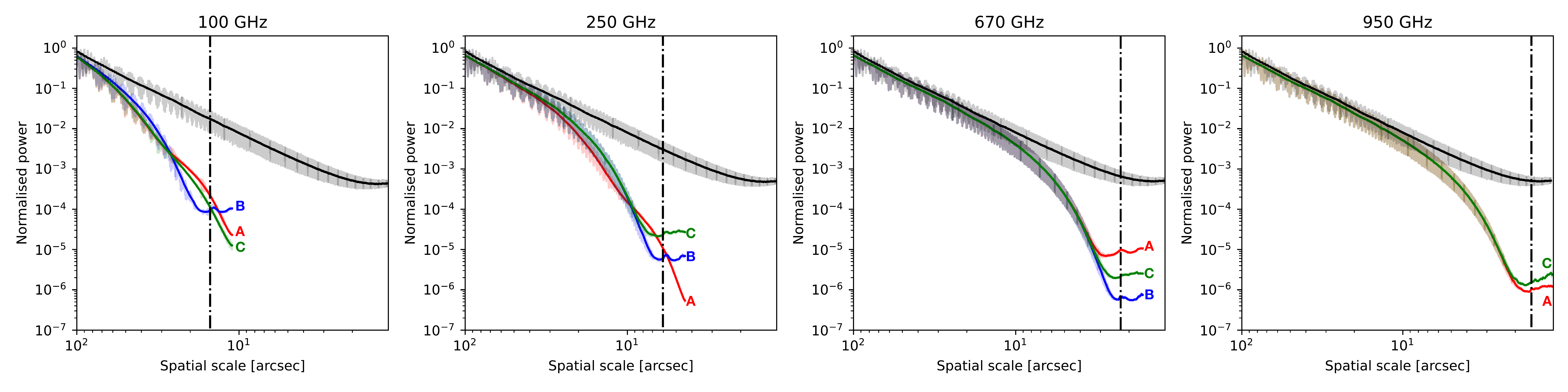}
        \label{fig:subfig_scan_A}
    \end{subfigure}
    \begin{subfigure}[b]{1\textwidth}
        \centering
        \includegraphics[width=\textwidth]{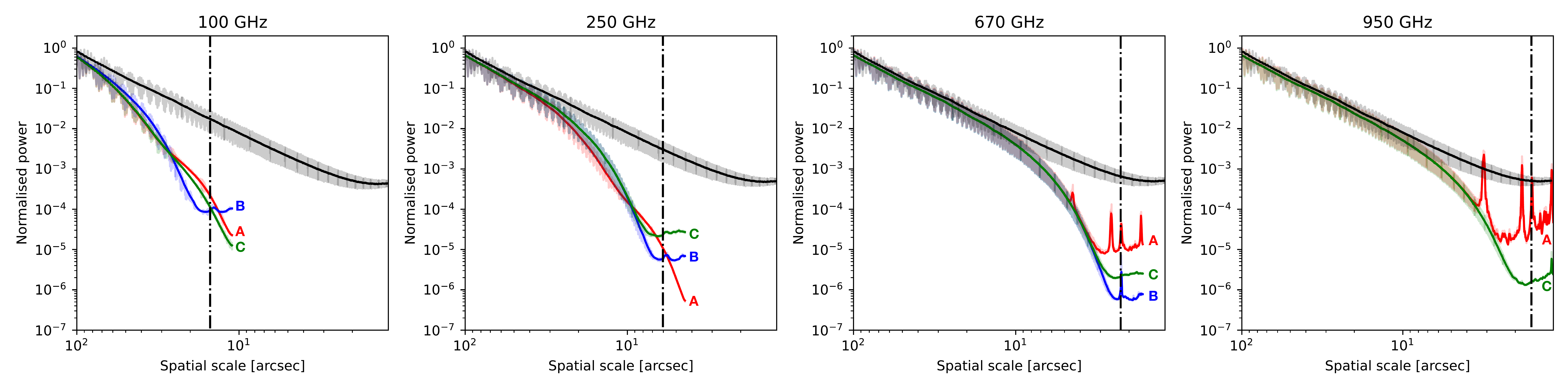}
        \label{fig:subfig_scan_B}
    \end{subfigure}
    \caption{
    Spatial power spectra for the full-disk maps at different frequencies (shown in different columns) for simulations with no imposed atmosphere (top row) and with a two-dimensional atmosphere (bottom row). The two-dimensional atmosphere is here a simplified approximation of atmospheric turbulence \citep[see][for a detailed overview]{2022PhRvD.105d2004M}. The \texttt{maria} code generates an atmosphere based on real weather data from the Chajnantor plateau \citep{2024OJAp....7E.118V}. The different \updated{instrumental setups} are represented by different colours: red for case A, blue for case B, and green for case C. For comparison, the power spectrum for the input map (light-grey shades) and the nominal angular resolution (dot-dashed vertical lines) are shown. The power spectra are overplotted with a smoothed line in a similar colour for clarity.}
    \label{fig:power_spectra_atmos}
\end{figure*}

In the top-row of Figure~\ref{fig:power_spectra_atmos}, the spatial power spectra are shown  for each of the three cases A, B, and C, for the available observing frequencies  (i.e. at 100, 250, 670, and 950 GHz for A and C, and for 100, 250, and 670\,GHz for Case B). 
See Figure~\ref{fig:scans_comp_fovs} for the corresponding input maps. 
The power spectra show greatly reduced power at smaller scales 
for all synthetic observations  
compared to the input map, which is observed across all frequencies and with all three cases.
The maps that required post-processing in the form of smoothing, i.e., the maps at 100 GHz for cases A and C, and the 250 GHz map for case C, show no \updated{sign} of a plateau at or near the spatial scale of beam at those frequencies. All other maps show a flattening of the power spectra at spatial scales just larger than the beam scale.

Having focused mainly on the time it takes to scan the solar disk, and not necessarily on the quality of the produced maps, we have up until now only considered idealised \texttt{maria} simulations with no atmosphere. However, the addition of an atmosphere is likely to impact the simulated observations, and therefore also the power spectra greatly. The three Cases A, B, and C are now rerun with a two-dimensional atmosphere imposed in \texttt{maria}. This atmosphere is based on real weather data collected at the Llano de Chajnantor, Chile, and should be representative for future AtLAST observations \citep{2022PhRvD.105d2004M}. The resulting power spectra are shown in Fig.~\ref{fig:power_spectra_atmos}. It is observed that the results for 100\,GHz and 250~GHz are seemingly not impacted by the imposed atmosphere, likely owing to the high atmospheric transmission at these frequencies  \citep{2009IEEEP..97.1463W}. However, at 670 and 950~GHz the atmosphere impacts the observations to a larger degree, increasing the spatial scales at which the power flattens out to the noise floor greatly. Interestingly, at 950~GHz the atmosphere seems to impact the Case A (small instrument and long scan) more than Case C (large instrument and short scan), with the noise becoming dominant at spatial scales of $\sim 3.5\arcsec$ with the Case A instrument, and at $\sim 2\arcsec$ with the Case C instrument. This would indicate that a large FOV instrument that uses less time to cover the solar disk would be preferable, as one could recover smaller scales due to the lesser impact of the atmosphere. To adequately study the effect of the atmosphere on AtLAST observations, one would have to take the time-domain into account by utilising an evolving map. One could thus study the atmosphere's effect on successfully capturing the smallest and most dynamic phenomena in the solar chromosphere. Such a study goes beyond the scope of this paper but should be carried in connection with instrumental design study. 

\section{Conclusions} \label{sec:conclusion}

The presented simulations of AtLAST observations of the Sun across a frequency range from 100 to 950~GHz show that the whole solar disk of the Sun can be scanned in under a minute, facilitating time series of full-disk scans with sub-minute cadence. This promising capability would be achievable even with instruments with detector element counts on the order of thousands, which is realistic already for a first-generation instrument for AtLAST. The anticipated full-disk map time series would be novel in terms of angular and temporal resolution for millimetre regime, clearly exceeding ALMA TP scans with $\sim 10$~minutes cadence and even at four times higher resolution. 

Our work demonstrates that the most suitable scan method of the solar disk greatly depends on the instrument in use, as instruments with large fields of view can support simpler scan patterns, and thus a higher resulting cadence, compared to the dense scan patterns required for instruments with a small instantaneous FOV. 
%Even a minute cadence would be a great improvement over current millimeter facilities that offer full-disk mapping at comparable resolutions \ct. 
Pushing the number of detector elements to the order of 100,000, a number that should be achievable at least for a second-generation instrument, we find scan times and thus cadences on the order of seconds to be achievable, here by utilising a simple circular scan pattern. Sequences of full-disk maps every few seconds would complement similar full-disk maps at higher frequencies, for example, those in the UV regime from the Solar Dynamics Observatory \citep{2012SoPh..275....3P, 2012SoPh..275...17L}, with a cadence of $12$ and $24$ seconds depending on the observed frequency. 

With a (sub-)minute cadence, AtLAST would be capable of addressing many of the science cases in solar millimetre astronomy. In combination with its large FOV, the high cadence would greatly increase the likelihood of AtLAST detecting flares. The large FOV would also be a great advantage in the study of millimetre radiation from solar prominences. Even at a lower resolution than ALMA, AtLAST would make an exemplary instrument for solar millimetre astronomy, both on its own and as a complement to ALMA observations, being able to study solar phenomena on larger spatial scales that exceed the limited FOV of an interferometric array such as ALMA.  

Based on our findings that high cadence full-disk solar observations with AtLAST are possible, we recommend that a dedicated instrumental study be performed for finding the optimal parameters of a solar- or at least solar-capable instrument for AtLAST, which would undoubtedly provide the solar millimetre community with extremely useful data. \updated{Such an instrument, although optimised for high-cadence full-disk scanning of the Sun could also be used for observing other extended sources or sky regions. In such cases, the scanning speed and time could be altered based on the required integration time to observe the source in question.}

\section*{Acknowledgements}
\updated{This project has received funding from the European Union's Horizon Europe and Horizon2020 research and innovation programmes under grant agreements No. 101188037 (AtLAST2) and No. 951815 (AtLAST). Views and opinions expressed are however those of the author(s) only and do not necessarily reflect those of the European Union or European Research Executive Agency. Neither the European Union nor the European Research Executive Agency can be held responsible for them.}
This work is supported by the Research Council of Norway through its Centres of Excellence scheme, project number 262622 (Rosseland Centre for Solar Physics). This paper makes use of the following ALMA data: ADS/JAO.ALMA\#2022.1.01544.S. ALMA is a partnership of ESO (representing its member states), NSF (USA) and NINS (Japan), together with NRC(Canada), MOST and ASIAA (Taiwan), and KASI (Republic of Korea), in co-operation with the Republic of Chile. 
The authors thank Evanthia Hatziminaoglou and Matt Griffin for their valuable comments. 

\newpage
\bibliography{main}

\begin{thebibliography}{39}%
\makeatletter
\providecommand \@ifxundefined [1]{%
 \@ifx{#1\undefined}
}%
\providecommand \@ifnum [1]{%
 \ifnum #1\expandafter \@firstoftwo
 \else \expandafter \@secondoftwo
 \fi
}%
\providecommand \@ifx [1]{%
 \ifx #1\expandafter \@firstoftwo
 \else \expandafter \@secondoftwo
 \fi
}%
\providecommand \natexlab [1]{#1}%
\providecommand \enquote  [1]{``#1''}%
\providecommand \bibnamefont  [1]{#1}%
\providecommand \bibfnamefont [1]{#1}%
\providecommand \citenamefont [1]{#1}%
\providecommand \href@noop [0]{\@secondoftwo}%
\providecommand \href [0]{\begingroup \@sanitize@url \@href}%
\providecommand \@href[1]{\@@startlink{#1}\@@href}%
\providecommand \@@href[1]{\endgroup#1\@@endlink}%
\providecommand \@sanitize@url [0]{\catcode `\\12\catcode `\$12\catcode `\&12\catcode `\#12\catcode `\^12\catcode `\_12\catcode `\%12\relax}%
\providecommand \@@startlink[1]{}%
\providecommand \@@endlink[0]{}%
\providecommand \url  [0]{\begingroup\@sanitize@url \@url }%
\providecommand \@url [1]{\endgroup\@href {#1}{\urlprefix }}%
\providecommand \urlprefix  [0]{URL }%
\providecommand \Eprint [0]{\href }%
\providecommand \doibase [0]{http://dx.doi.org/}%
\providecommand \selectlanguage [0]{\@gobble}%
\providecommand \bibinfo  [0]{\@secondoftwo}%
\providecommand \bibfield  [0]{\@secondoftwo}%
\providecommand \translation [1]{[#1]}%
\providecommand \BibitemOpen [0]{}%
\providecommand \bibitemStop [0]{}%
\providecommand \bibitemNoStop [0]{.\EOS\space}%
\providecommand \EOS [0]{\spacefactor3000\relax}%
\providecommand \BibitemShut  [1]{\csname bibitem#1\endcsname}%
\let\auto@bib@innerbib\@empty
%</preamble>
\bibitem [{\citenamefont {{Wedemeyer}}\ \emph {et~al.}(2016)\citenamefont {{Wedemeyer}}, \citenamefont {{Bastian}}, \citenamefont {{Braj{\v{s}}a}}, \citenamefont {{Hudson}}, \citenamefont {{Fleishman}}, \citenamefont {{Loukitcheva}}, \citenamefont {{Fleck}}, \citenamefont {{Kontar}}, \citenamefont {{De Pontieu}}, \citenamefont {{Yagoubov}}, \citenamefont {{Tiwari}}, \citenamefont {{Soler}}, \citenamefont {{Black}}, \citenamefont {{Antolin}}, \citenamefont {{Scullion}}, \citenamefont {{Gun{\'a}r}}, \citenamefont {{Labrosse}}, \citenamefont {{Ludwig}}, \citenamefont {{Benz}}, \citenamefont {{White}}, \citenamefont {{Hauschildt}}, \citenamefont {{Doyle}}, \citenamefont {{Nakariakov}}, \citenamefont {{Ayres}}, \citenamefont {{Heinzel}}, \citenamefont {{Karlicky}}, \citenamefont {{Van Doorsselaere}}, \citenamefont {{Gary}}, \citenamefont {{Alissandrakis}}, \citenamefont {{Nindos}}, \citenamefont {{Solanki}}, \citenamefont {{Rouppe van der Voort}}, \citenamefont {{Shimojo}}, \citenamefont {{Kato}}, \citenamefont
  {{Zaqarashvili}}, \citenamefont {{Perez}}, \citenamefont {{Selhorst}},\ and\ \citenamefont {{Barta}}}]{2016SSRv..200....1W}%
  \BibitemOpen
  \bibfield  {author} {\bibinfo {author} {\bibfnamefont {S.}~\bibnamefont {{Wedemeyer}}}, \bibinfo {author} {\bibfnamefont {T.}~\bibnamefont {{Bastian}}}, \bibinfo {author} {\bibfnamefont {R.}~\bibnamefont {{Braj{\v{s}}a}}}, \bibinfo {author} {\bibfnamefont {H.}~\bibnamefont {{Hudson}}}, \bibinfo {author} {\bibfnamefont {G.}~\bibnamefont {{Fleishman}}}, \bibinfo {author} {\bibfnamefont {M.}~\bibnamefont {{Loukitcheva}}}, \bibinfo {author} {\bibfnamefont {B.}~\bibnamefont {{Fleck}}}, \bibinfo {author} {\bibfnamefont {E.~P.}\ \bibnamefont {{Kontar}}}, \bibinfo {author} {\bibfnamefont {B.}~\bibnamefont {{De Pontieu}}}, \bibinfo {author} {\bibfnamefont {P.}~\bibnamefont {{Yagoubov}}}, \bibinfo {author} {\bibfnamefont {S.~K.}\ \bibnamefont {{Tiwari}}}, \bibinfo {author} {\bibfnamefont {R.}~\bibnamefont {{Soler}}}, \bibinfo {author} {\bibfnamefont {J.~H.}\ \bibnamefont {{Black}}}, \bibinfo {author} {\bibfnamefont {P.}~\bibnamefont {{Antolin}}}, \bibinfo {author} {\bibfnamefont {E.}~\bibnamefont {{Scullion}}},
  \bibinfo {author} {\bibfnamefont {S.}~\bibnamefont {{Gun{\'a}r}}}, \bibinfo {author} {\bibfnamefont {N.}~\bibnamefont {{Labrosse}}}, \bibinfo {author} {\bibfnamefont {H.~G.}\ \bibnamefont {{Ludwig}}}, \bibinfo {author} {\bibfnamefont {A.~O.}\ \bibnamefont {{Benz}}}, \bibinfo {author} {\bibfnamefont {S.~M.}\ \bibnamefont {{White}}}, \bibinfo {author} {\bibfnamefont {P.}~\bibnamefont {{Hauschildt}}}, \bibinfo {author} {\bibfnamefont {J.~G.}\ \bibnamefont {{Doyle}}}, \bibinfo {author} {\bibfnamefont {V.~M.}\ \bibnamefont {{Nakariakov}}}, \bibinfo {author} {\bibfnamefont {T.}~\bibnamefont {{Ayres}}}, \bibinfo {author} {\bibfnamefont {P.}~\bibnamefont {{Heinzel}}}, \bibinfo {author} {\bibfnamefont {M.}~\bibnamefont {{Karlicky}}}, \bibinfo {author} {\bibfnamefont {T.}~\bibnamefont {{Van Doorsselaere}}}, \bibinfo {author} {\bibfnamefont {D.}~\bibnamefont {{Gary}}}, \bibinfo {author} {\bibfnamefont {C.~E.}\ \bibnamefont {{Alissandrakis}}}, \bibinfo {author} {\bibfnamefont {A.}~\bibnamefont {{Nindos}}}, \bibinfo
  {author} {\bibfnamefont {S.~K.}\ \bibnamefont {{Solanki}}}, \bibinfo {author} {\bibfnamefont {L.}~\bibnamefont {{Rouppe van der Voort}}}, \bibinfo {author} {\bibfnamefont {M.}~\bibnamefont {{Shimojo}}}, \bibinfo {author} {\bibfnamefont {Y.}~\bibnamefont {{Kato}}}, \bibinfo {author} {\bibfnamefont {T.}~\bibnamefont {{Zaqarashvili}}}, \bibinfo {author} {\bibfnamefont {E.}~\bibnamefont {{Perez}}}, \bibinfo {author} {\bibfnamefont {C.~L.}\ \bibnamefont {{Selhorst}}}, \ and\ \bibinfo {author} {\bibfnamefont {M.}~\bibnamefont {{Barta}}},\ }\href {\doibase 10.1007/s11214-015-0229-9} {\bibfield  {journal} {\bibinfo  {journal} {\ssr}\ }\textbf {\bibinfo {volume} {200}},\ \bibinfo {pages} {1} (\bibinfo {year} {2016})},\ \Eprint {http://arxiv.org/abs/1504.06887} {arXiv:1504.06887 [astro-ph.SR]} \BibitemShut {NoStop}%
\bibitem [{\citenamefont {{Rouppe van der Voort}}\ \emph {et~al.}(2020)\citenamefont {{Rouppe van der Voort}}, \citenamefont {{De Pontieu}}, \citenamefont {{Carlsson}}, \citenamefont {{de la Cruz Rodr{\'\i}guez}}, \citenamefont {{Bose}}, \citenamefont {{Chintzoglou}}, \citenamefont {{Drews}}, \citenamefont {{Froment}}, \citenamefont {{Go{\v{s}}i{\'c}}}, \citenamefont {{Graham}}, \citenamefont {{Hansteen}}, \citenamefont {{Henriques}}, \citenamefont {{Jafarzadeh}}, \citenamefont {{Joshi}}, \citenamefont {{Kleint}}, \citenamefont {{Kohutova}}, \citenamefont {{Leifsen}}, \citenamefont {{Mart{\'\i}nez-Sykora}}, \citenamefont {{N{\'o}brega-Siverio}}, \citenamefont {{Ortiz}}, \citenamefont {{Pereira}}, \citenamefont {{Popovas}}, \citenamefont {{Quintero Noda}}, \citenamefont {{Sainz Dalda}}, \citenamefont {{Scharmer}}, \citenamefont {{Schmit}}, \citenamefont {{Scullion}}, \citenamefont {{Skogsrud}}, \citenamefont {{Szydlarski}}, \citenamefont {{Timmons}}, \citenamefont {{Vissers}}, \citenamefont {{Woods}},\ and\
  \citenamefont {{Zacharias}}}]{2020A&A...641A.146R}%
  \BibitemOpen
  \bibfield  {author} {\bibinfo {author} {\bibfnamefont {L.~H.~M.}\ \bibnamefont {{Rouppe van der Voort}}}, \bibinfo {author} {\bibfnamefont {B.}~\bibnamefont {{De Pontieu}}}, \bibinfo {author} {\bibfnamefont {M.}~\bibnamefont {{Carlsson}}}, \bibinfo {author} {\bibfnamefont {J.}~\bibnamefont {{de la Cruz Rodr{\'\i}guez}}}, \bibinfo {author} {\bibfnamefont {S.}~\bibnamefont {{Bose}}}, \bibinfo {author} {\bibfnamefont {G.}~\bibnamefont {{Chintzoglou}}}, \bibinfo {author} {\bibfnamefont {A.}~\bibnamefont {{Drews}}}, \bibinfo {author} {\bibfnamefont {C.}~\bibnamefont {{Froment}}}, \bibinfo {author} {\bibfnamefont {M.}~\bibnamefont {{Go{\v{s}}i{\'c}}}}, \bibinfo {author} {\bibfnamefont {D.~R.}\ \bibnamefont {{Graham}}}, \bibinfo {author} {\bibfnamefont {V.~H.}\ \bibnamefont {{Hansteen}}}, \bibinfo {author} {\bibfnamefont {V.~M.~J.}\ \bibnamefont {{Henriques}}}, \bibinfo {author} {\bibfnamefont {S.}~\bibnamefont {{Jafarzadeh}}}, \bibinfo {author} {\bibfnamefont {J.}~\bibnamefont {{Joshi}}}, \bibinfo {author}
  {\bibfnamefont {L.}~\bibnamefont {{Kleint}}}, \bibinfo {author} {\bibfnamefont {P.}~\bibnamefont {{Kohutova}}}, \bibinfo {author} {\bibfnamefont {T.}~\bibnamefont {{Leifsen}}}, \bibinfo {author} {\bibfnamefont {J.}~\bibnamefont {{Mart{\'\i}nez-Sykora}}}, \bibinfo {author} {\bibfnamefont {D.}~\bibnamefont {{N{\'o}brega-Siverio}}}, \bibinfo {author} {\bibfnamefont {A.}~\bibnamefont {{Ortiz}}}, \bibinfo {author} {\bibfnamefont {T.~M.~D.}\ \bibnamefont {{Pereira}}}, \bibinfo {author} {\bibfnamefont {A.}~\bibnamefont {{Popovas}}}, \bibinfo {author} {\bibfnamefont {C.}~\bibnamefont {{Quintero Noda}}}, \bibinfo {author} {\bibfnamefont {A.}~\bibnamefont {{Sainz Dalda}}}, \bibinfo {author} {\bibfnamefont {G.~B.}\ \bibnamefont {{Scharmer}}}, \bibinfo {author} {\bibfnamefont {D.}~\bibnamefont {{Schmit}}}, \bibinfo {author} {\bibfnamefont {E.}~\bibnamefont {{Scullion}}}, \bibinfo {author} {\bibfnamefont {H.}~\bibnamefont {{Skogsrud}}}, \bibinfo {author} {\bibfnamefont {M.}~\bibnamefont {{Szydlarski}}}, \bibinfo
  {author} {\bibfnamefont {R.}~\bibnamefont {{Timmons}}}, \bibinfo {author} {\bibfnamefont {G.~J.~M.}\ \bibnamefont {{Vissers}}}, \bibinfo {author} {\bibfnamefont {M.~M.}\ \bibnamefont {{Woods}}}, \ and\ \bibinfo {author} {\bibfnamefont {P.}~\bibnamefont {{Zacharias}}},\ }\href {\doibase 10.1051/0004-6361/202038732} {\bibfield  {journal} {\bibinfo  {journal} {\aap}\ }\textbf {\bibinfo {volume} {641}},\ \bibinfo {eid} {A146} (\bibinfo {year} {2020})},\ \Eprint {http://arxiv.org/abs/2005.14175} {arXiv:2005.14175 [astro-ph.SR]} \BibitemShut {NoStop}%
\bibitem [{\citenamefont {{Thoen Faber}}\ \emph {et~al.}(2025)\citenamefont {{Thoen Faber}}, \citenamefont {{Joshi}}, \citenamefont {{van der Voort}}, \citenamefont {{Wedemeyer}}, \citenamefont {{Fletcher}}, \citenamefont {{Aulanier}},\ and\ \citenamefont {{N{\'o}brega-Siverio}}}]{2025A&A...693A...8T}%
  \BibitemOpen
  \bibfield  {author} {\bibinfo {author} {\bibfnamefont {J.}~\bibnamefont {{Thoen Faber}}}, \bibinfo {author} {\bibfnamefont {R.}~\bibnamefont {{Joshi}}}, \bibinfo {author} {\bibfnamefont {L.~R.}\ \bibnamefont {{van der Voort}}}, \bibinfo {author} {\bibfnamefont {S.}~\bibnamefont {{Wedemeyer}}}, \bibinfo {author} {\bibfnamefont {L.}~\bibnamefont {{Fletcher}}}, \bibinfo {author} {\bibfnamefont {G.}~\bibnamefont {{Aulanier}}}, \ and\ \bibinfo {author} {\bibfnamefont {D.}~\bibnamefont {{N{\'o}brega-Siverio}}},\ }\href {\doibase 10.1051/0004-6361/202452370} {\bibfield  {journal} {\bibinfo  {journal} {\aap}\ }\textbf {\bibinfo {volume} {693}},\ \bibinfo {eid} {A8} (\bibinfo {year} {2025})},\ \Eprint {http://arxiv.org/abs/2411.18233} {arXiv:2411.18233 [astro-ph.SR]} \BibitemShut {NoStop}%
\bibitem [{\citenamefont {{de la Cruz Rodr{\'\i}guez}}\ \emph {et~al.}(2016)\citenamefont {{de la Cruz Rodr{\'\i}guez}}, \citenamefont {{Leenaarts}},\ and\ \citenamefont {{Asensio Ramos}}}]{2016ApJ...830L..30D}%
  \BibitemOpen
  \bibfield  {author} {\bibinfo {author} {\bibfnamefont {J.}~\bibnamefont {{de la Cruz Rodr{\'\i}guez}}}, \bibinfo {author} {\bibfnamefont {J.}~\bibnamefont {{Leenaarts}}}, \ and\ \bibinfo {author} {\bibfnamefont {A.}~\bibnamefont {{Asensio Ramos}}},\ }\href {\doibase 10.3847/2041-8205/830/2/L30} {\bibfield  {journal} {\bibinfo  {journal} {\apjl}\ }\textbf {\bibinfo {volume} {830}},\ \bibinfo {eid} {L30} (\bibinfo {year} {2016})},\ \Eprint {http://arxiv.org/abs/1609.09527} {arXiv:1609.09527 [astro-ph.SR]} \BibitemShut {NoStop}%
\bibitem [{\citenamefont {{Bastian}}\ \emph {et~al.}(2018)\citenamefont {{Bastian}}, \citenamefont {{B{\'a}rta}}, \citenamefont {{Braj{\v{s}}a}}, \citenamefont {{Chen}}, \citenamefont {{Pontieu}}, \citenamefont {{Gary}}, \citenamefont {{Fleishman}}, \citenamefont {{Hales}}, \citenamefont {{Iwai}}, \citenamefont {{Hudson}}, \citenamefont {{Kim}}, \citenamefont {{Kobelski}}, \citenamefont {{Loukitcheva}}, \citenamefont {{Shimojo}}, \citenamefont {{Skoki{\'c}}}, \citenamefont {{Wedemeyer}}, \citenamefont {{White}},\ and\ \citenamefont {{Yan}}}]{2018Msngr.171...25B}%
  \BibitemOpen
  \bibfield  {author} {\bibinfo {author} {\bibfnamefont {T.~S.}\ \bibnamefont {{Bastian}}}, \bibinfo {author} {\bibfnamefont {M.}~\bibnamefont {{B{\'a}rta}}}, \bibinfo {author} {\bibfnamefont {R.}~\bibnamefont {{Braj{\v{s}}a}}}, \bibinfo {author} {\bibfnamefont {B.}~\bibnamefont {{Chen}}}, \bibinfo {author} {\bibfnamefont {B.~D.}\ \bibnamefont {{Pontieu}}}, \bibinfo {author} {\bibfnamefont {D.~E.}\ \bibnamefont {{Gary}}}, \bibinfo {author} {\bibfnamefont {G.~D.}\ \bibnamefont {{Fleishman}}}, \bibinfo {author} {\bibfnamefont {A.~S.}\ \bibnamefont {{Hales}}}, \bibinfo {author} {\bibfnamefont {K.}~\bibnamefont {{Iwai}}}, \bibinfo {author} {\bibfnamefont {H.}~\bibnamefont {{Hudson}}}, \bibinfo {author} {\bibfnamefont {S.}~\bibnamefont {{Kim}}}, \bibinfo {author} {\bibfnamefont {A.}~\bibnamefont {{Kobelski}}}, \bibinfo {author} {\bibfnamefont {M.}~\bibnamefont {{Loukitcheva}}}, \bibinfo {author} {\bibfnamefont {M.}~\bibnamefont {{Shimojo}}}, \bibinfo {author} {\bibfnamefont {I.}~\bibnamefont {{Skoki{\'c}}}},
  \bibinfo {author} {\bibfnamefont {S.}~\bibnamefont {{Wedemeyer}}}, \bibinfo {author} {\bibfnamefont {S.~M.}\ \bibnamefont {{White}}}, \ and\ \bibinfo {author} {\bibfnamefont {Y.}~\bibnamefont {{Yan}}},\ }\href@noop {} {\bibfield  {journal} {\bibinfo  {journal} {The Messenger}\ }\textbf {\bibinfo {volume} {171}},\ \bibinfo {pages} {25} (\bibinfo {year} {2018})}\BibitemShut {NoStop}%
\bibitem [{\citenamefont {{Bastian}}(2002)}]{2002AN....323..271B}%
  \BibitemOpen
  \bibfield  {author} {\bibinfo {author} {\bibfnamefont {T.~S.}\ \bibnamefont {{Bastian}}},\ }\href {\doibase 10.1002/1521-3994(200208)323:3/4<271::AID-ASNA271>3.0.CO;2-1} {\bibfield  {journal} {\bibinfo  {journal} {Astronomische Nachrichten}\ }\textbf {\bibinfo {volume} {323}},\ \bibinfo {pages} {271} (\bibinfo {year} {2002})}\BibitemShut {NoStop}%
\bibitem [{\citenamefont {{Wedemeyer}}\ \emph {et~al.}(2024)\citenamefont {{Wedemeyer}}, \citenamefont {{Barta}}, \citenamefont {{Braj{\v{s}}a}}, \citenamefont {{Chai}}, \citenamefont {{Costa}}, \citenamefont {{Gary}}, \citenamefont {{Gimenez de Castro}}, \citenamefont {{Gunar}}, \citenamefont {{Fleishman}}, \citenamefont {{Hales}}, \citenamefont {{Hudson}}, \citenamefont {{Kirkaune}}, \citenamefont {{Mohan}}, \citenamefont {{Motorina}}, \citenamefont {{Pellizzoni}}, \citenamefont {{Saberi}}, \citenamefont {{Selhorst}}, \citenamefont {{Simoes}}, \citenamefont {{Shimojo}}, \citenamefont {{Skoki{\'c}}}, \citenamefont {{Sudar}}, \citenamefont {{Menezes}}, \citenamefont {{White}}, \citenamefont {{Booth}}, \citenamefont {{Klaassen}}, \citenamefont {{Cicone}}, \citenamefont {{Mroczkowski}}, \citenamefont {{Cordiner}}, \citenamefont {{Di Mascolo}}, \citenamefont {{Johnstone}}, \citenamefont {{van Kampen}}, \citenamefont {{Lee}}, \citenamefont {{Liu}}, \citenamefont {{Maccarone}}, \citenamefont {{Orlowski-Scherer}},
  \citenamefont {{Saintonge}}, \citenamefont {{Smith}},\ and\ \citenamefont {{Thelen}}}]{2024ORE.....4..140W}%
  \BibitemOpen
  \bibfield  {author} {\bibinfo {author} {\bibfnamefont {S.}~\bibnamefont {{Wedemeyer}}}, \bibinfo {author} {\bibfnamefont {M.}~\bibnamefont {{Barta}}}, \bibinfo {author} {\bibfnamefont {R.}~\bibnamefont {{Braj{\v{s}}a}}}, \bibinfo {author} {\bibfnamefont {Y.}~\bibnamefont {{Chai}}}, \bibinfo {author} {\bibfnamefont {J.}~\bibnamefont {{Costa}}}, \bibinfo {author} {\bibfnamefont {D.}~\bibnamefont {{Gary}}}, \bibinfo {author} {\bibfnamefont {G.}~\bibnamefont {{Gimenez de Castro}}}, \bibinfo {author} {\bibfnamefont {S.}~\bibnamefont {{Gunar}}}, \bibinfo {author} {\bibfnamefont {G.}~\bibnamefont {{Fleishman}}}, \bibinfo {author} {\bibfnamefont {A.}~\bibnamefont {{Hales}}}, \bibinfo {author} {\bibfnamefont {H.}~\bibnamefont {{Hudson}}}, \bibinfo {author} {\bibfnamefont {M.}~\bibnamefont {{Kirkaune}}}, \bibinfo {author} {\bibfnamefont {A.}~\bibnamefont {{Mohan}}}, \bibinfo {author} {\bibfnamefont {G.}~\bibnamefont {{Motorina}}}, \bibinfo {author} {\bibfnamefont {A.}~\bibnamefont {{Pellizzoni}}}, \bibinfo {author}
  {\bibfnamefont {M.}~\bibnamefont {{Saberi}}}, \bibinfo {author} {\bibfnamefont {C.~L.}\ \bibnamefont {{Selhorst}}}, \bibinfo {author} {\bibfnamefont {P.~J.~A.}\ \bibnamefont {{Simoes}}}, \bibinfo {author} {\bibfnamefont {M.}~\bibnamefont {{Shimojo}}}, \bibinfo {author} {\bibfnamefont {I.}~\bibnamefont {{Skoki{\'c}}}}, \bibinfo {author} {\bibfnamefont {D.}~\bibnamefont {{Sudar}}}, \bibinfo {author} {\bibfnamefont {F.}~\bibnamefont {{Menezes}}}, \bibinfo {author} {\bibfnamefont {S.~M.}\ \bibnamefont {{White}}}, \bibinfo {author} {\bibfnamefont {M.}~\bibnamefont {{Booth}}}, \bibinfo {author} {\bibfnamefont {P.}~\bibnamefont {{Klaassen}}}, \bibinfo {author} {\bibfnamefont {C.}~\bibnamefont {{Cicone}}}, \bibinfo {author} {\bibfnamefont {T.}~\bibnamefont {{Mroczkowski}}}, \bibinfo {author} {\bibfnamefont {M.~A.}\ \bibnamefont {{Cordiner}}}, \bibinfo {author} {\bibfnamefont {L.}~\bibnamefont {{Di Mascolo}}}, \bibinfo {author} {\bibfnamefont {D.}~\bibnamefont {{Johnstone}}}, \bibinfo {author} {\bibfnamefont
  {E.}~\bibnamefont {{van Kampen}}}, \bibinfo {author} {\bibfnamefont {M.}~\bibnamefont {{Lee}}}, \bibinfo {author} {\bibfnamefont {D.}~\bibnamefont {{Liu}}}, \bibinfo {author} {\bibfnamefont {T.}~\bibnamefont {{Maccarone}}}, \bibinfo {author} {\bibfnamefont {J.}~\bibnamefont {{Orlowski-Scherer}}}, \bibinfo {author} {\bibfnamefont {A.}~\bibnamefont {{Saintonge}}}, \bibinfo {author} {\bibfnamefont {M.}~\bibnamefont {{Smith}}}, \ and\ \bibinfo {author} {\bibfnamefont {A.~E.}\ \bibnamefont {{Thelen}}},\ }\href {\doibase 10.12688/openreseurope.17453.1} {\bibfield  {journal} {\bibinfo  {journal} {Open Research Europe}\ }\textbf {\bibinfo {volume} {4}},\ \bibinfo {eid} {140} (\bibinfo {year} {2024})},\ \Eprint {http://arxiv.org/abs/2403.00920} {arXiv:2403.00920 [astro-ph.SR]} \BibitemShut {NoStop}%
\bibitem [{\citenamefont {{Wootten}}\ and\ \citenamefont {{Thompson}}(2009)}]{2009IEEEP..97.1463W}%
  \BibitemOpen
  \bibfield  {author} {\bibinfo {author} {\bibfnamefont {A.}~\bibnamefont {{Wootten}}}\ and\ \bibinfo {author} {\bibfnamefont {A.~R.}\ \bibnamefont {{Thompson}}},\ }\href {\doibase 10.1109/JPROC.2009.2020572} {\bibfield  {journal} {\bibinfo  {journal} {IEEE Proceedings}\ }\textbf {\bibinfo {volume} {97}},\ \bibinfo {pages} {1463} (\bibinfo {year} {2009})},\ \Eprint {http://arxiv.org/abs/0904.3739} {arXiv:0904.3739 [astro-ph.IM]} \BibitemShut {NoStop}%
\bibitem [{\citenamefont {{Bastian}}\ \emph {et~al.}(2022)\citenamefont {{Bastian}}, \citenamefont {{Shimojo}}, \citenamefont {{B{\'a}rta}}, \citenamefont {{White}},\ and\ \citenamefont {{Iwai}}}]{2022FrASS...9.7368B}%
  \BibitemOpen
  \bibfield  {author} {\bibinfo {author} {\bibfnamefont {T.~S.}\ \bibnamefont {{Bastian}}}, \bibinfo {author} {\bibfnamefont {M.}~\bibnamefont {{Shimojo}}}, \bibinfo {author} {\bibfnamefont {M.}~\bibnamefont {{B{\'a}rta}}}, \bibinfo {author} {\bibfnamefont {S.~M.}\ \bibnamefont {{White}}}, \ and\ \bibinfo {author} {\bibfnamefont {K.}~\bibnamefont {{Iwai}}},\ }\href {\doibase 10.3389/fspas.2022.977368} {\bibfield  {journal} {\bibinfo  {journal} {Frontiers in Astronomy and Space Sciences}\ }\textbf {\bibinfo {volume} {9}},\ \bibinfo {eid} {977368} (\bibinfo {year} {2022})},\ \Eprint {http://arxiv.org/abs/2209.01659} {arXiv:2209.01659 [astro-ph.IM]} \BibitemShut {NoStop}%
\bibitem [{\citenamefont {{Shimizu}}\ \emph {et~al.}(2021)\citenamefont {{Shimizu}}, \citenamefont {{Shimojo}},\ and\ \citenamefont {{Abe}}}]{2021ApJ...922..113S}%
  \BibitemOpen
  \bibfield  {author} {\bibinfo {author} {\bibfnamefont {T.}~\bibnamefont {{Shimizu}}}, \bibinfo {author} {\bibfnamefont {M.}~\bibnamefont {{Shimojo}}}, \ and\ \bibinfo {author} {\bibfnamefont {M.}~\bibnamefont {{Abe}}},\ }\href {\doibase 10.3847/1538-4357/ac27a4} {\bibfield  {journal} {\bibinfo  {journal} {\apj}\ }\textbf {\bibinfo {volume} {922}},\ \bibinfo {eid} {113} (\bibinfo {year} {2021})},\ \Eprint {http://arxiv.org/abs/2109.11215} {arXiv:2109.11215 [astro-ph.SR]} \BibitemShut {NoStop}%
\bibitem [{\citenamefont {{Skoki{\'c}}}\ \emph {et~al.}(2023)\citenamefont {{Skoki{\'c}}}, \citenamefont {{Benz}}, \citenamefont {{Braj{\v{s}}a}}, \citenamefont {{Sudar}}, \citenamefont {{Matkovi{\'c}}},\ and\ \citenamefont {{B{\'a}rta}}}]{2023A&A...669A.156S}%
  \BibitemOpen
  \bibfield  {author} {\bibinfo {author} {\bibfnamefont {I.}~\bibnamefont {{Skoki{\'c}}}}, \bibinfo {author} {\bibfnamefont {A.~O.}\ \bibnamefont {{Benz}}}, \bibinfo {author} {\bibfnamefont {R.}~\bibnamefont {{Braj{\v{s}}a}}}, \bibinfo {author} {\bibfnamefont {D.}~\bibnamefont {{Sudar}}}, \bibinfo {author} {\bibfnamefont {F.}~\bibnamefont {{Matkovi{\'c}}}}, \ and\ \bibinfo {author} {\bibfnamefont {M.}~\bibnamefont {{B{\'a}rta}}},\ }\href {\doibase 10.1051/0004-6361/202244532} {\bibfield  {journal} {\bibinfo  {journal} {\aap}\ }\textbf {\bibinfo {volume} {669}},\ \bibinfo {eid} {A156} (\bibinfo {year} {2023})},\ \Eprint {http://arxiv.org/abs/2211.16935} {arXiv:2211.16935 [astro-ph.SR]} \BibitemShut {NoStop}%
\bibitem [{\citenamefont {{White}}\ \emph {et~al.}(2017)\citenamefont {{White}}, \citenamefont {{Iwai}}, \citenamefont {{Phillips}}, \citenamefont {{Hills}}, \citenamefont {{Hirota}}, \citenamefont {{Yagoubov}}, \citenamefont {{Siringo}}, \citenamefont {{Shimojo}}, \citenamefont {{Bastian}}, \citenamefont {{Hales}}, \citenamefont {{Sawada}}, \citenamefont {{Asayama}}, \citenamefont {{Sugimoto}}, \citenamefont {{Marson}}, \citenamefont {{Kawasaki}}, \citenamefont {{Muller}}, \citenamefont {{Nakazato}}, \citenamefont {{Sugimoto}}, \citenamefont {{Braj{\v{s}}a}}, \citenamefont {{Skoki{\'c}}}, \citenamefont {{B{\'a}rta}}, \citenamefont {{Kim}}, \citenamefont {{Remijan}}, \citenamefont {{de Gregorio}}, \citenamefont {{Corder}}, \citenamefont {{Hudson}}, \citenamefont {{Loukitcheva}}, \citenamefont {{Chen}}, \citenamefont {{De Pontieu}}, \citenamefont {{Fleishmann}}, \citenamefont {{Gary}}, \citenamefont {{Kobelski}}, \citenamefont {{Wedemeyer}},\ and\ \citenamefont {{Yan}}}]{2017SoPh..292...88W}%
  \BibitemOpen
  \bibfield  {author} {\bibinfo {author} {\bibfnamefont {S.~M.}\ \bibnamefont {{White}}}, \bibinfo {author} {\bibfnamefont {K.}~\bibnamefont {{Iwai}}}, \bibinfo {author} {\bibfnamefont {N.~M.}\ \bibnamefont {{Phillips}}}, \bibinfo {author} {\bibfnamefont {R.~E.}\ \bibnamefont {{Hills}}}, \bibinfo {author} {\bibfnamefont {A.}~\bibnamefont {{Hirota}}}, \bibinfo {author} {\bibfnamefont {P.}~\bibnamefont {{Yagoubov}}}, \bibinfo {author} {\bibfnamefont {G.}~\bibnamefont {{Siringo}}}, \bibinfo {author} {\bibfnamefont {M.}~\bibnamefont {{Shimojo}}}, \bibinfo {author} {\bibfnamefont {T.~S.}\ \bibnamefont {{Bastian}}}, \bibinfo {author} {\bibfnamefont {A.~S.}\ \bibnamefont {{Hales}}}, \bibinfo {author} {\bibfnamefont {T.}~\bibnamefont {{Sawada}}}, \bibinfo {author} {\bibfnamefont {S.}~\bibnamefont {{Asayama}}}, \bibinfo {author} {\bibfnamefont {M.}~\bibnamefont {{Sugimoto}}}, \bibinfo {author} {\bibfnamefont {R.~G.}\ \bibnamefont {{Marson}}}, \bibinfo {author} {\bibfnamefont {W.}~\bibnamefont {{Kawasaki}}}, \bibinfo
  {author} {\bibfnamefont {E.}~\bibnamefont {{Muller}}}, \bibinfo {author} {\bibfnamefont {T.}~\bibnamefont {{Nakazato}}}, \bibinfo {author} {\bibfnamefont {K.}~\bibnamefont {{Sugimoto}}}, \bibinfo {author} {\bibfnamefont {R.}~\bibnamefont {{Braj{\v{s}}a}}}, \bibinfo {author} {\bibfnamefont {I.}~\bibnamefont {{Skoki{\'c}}}}, \bibinfo {author} {\bibfnamefont {M.}~\bibnamefont {{B{\'a}rta}}}, \bibinfo {author} {\bibfnamefont {S.}~\bibnamefont {{Kim}}}, \bibinfo {author} {\bibfnamefont {A.~J.}\ \bibnamefont {{Remijan}}}, \bibinfo {author} {\bibfnamefont {I.}~\bibnamefont {{de Gregorio}}}, \bibinfo {author} {\bibfnamefont {S.~A.}\ \bibnamefont {{Corder}}}, \bibinfo {author} {\bibfnamefont {H.~S.}\ \bibnamefont {{Hudson}}}, \bibinfo {author} {\bibfnamefont {M.}~\bibnamefont {{Loukitcheva}}}, \bibinfo {author} {\bibfnamefont {B.}~\bibnamefont {{Chen}}}, \bibinfo {author} {\bibfnamefont {B.}~\bibnamefont {{De Pontieu}}}, \bibinfo {author} {\bibfnamefont {G.~D.}\ \bibnamefont {{Fleishmann}}}, \bibinfo {author}
  {\bibfnamefont {D.~E.}\ \bibnamefont {{Gary}}}, \bibinfo {author} {\bibfnamefont {A.}~\bibnamefont {{Kobelski}}}, \bibinfo {author} {\bibfnamefont {S.}~\bibnamefont {{Wedemeyer}}}, \ and\ \bibinfo {author} {\bibfnamefont {Y.}~\bibnamefont {{Yan}}},\ }\href {\doibase 10.1007/s11207-017-1123-2} {\bibfield  {journal} {\bibinfo  {journal} {\solphys}\ }\textbf {\bibinfo {volume} {292}},\ \bibinfo {eid} {88} (\bibinfo {year} {2017})},\ \Eprint {http://arxiv.org/abs/1705.04766} {arXiv:1705.04766 [astro-ph.SR]} \BibitemShut {NoStop}%
\bibitem [{\citenamefont {{Mroczkowski}}\ \emph {et~al.}(2025)\citenamefont {{Mroczkowski}}, \citenamefont {{Gallardo}}, \citenamefont {{Timpe}}, \citenamefont {{Kiselev}}, \citenamefont {{Groh}}, \citenamefont {{Kaercher}}, \citenamefont {{Reichert}}, \citenamefont {{Cicone}}, \citenamefont {{Puddu}}, \citenamefont {{Dubois-dit-Bonclaude}}, \citenamefont {{Bok}}, \citenamefont {{Dahl}}, \citenamefont {{Macintosh}}, \citenamefont {{Dicker}}, \citenamefont {{Viole}}, \citenamefont {{Sartori}}, \citenamefont {{Andr{\'e}s Valenzuela Venegas}}, \citenamefont {{Zeyringer}}, \citenamefont {{Niemack}}, \citenamefont {{Poppi}}, \citenamefont {{Olguin}}, \citenamefont {{Hatziminaoglou}}, \citenamefont {{De Breuck}}, \citenamefont {{Klaassen}}, \citenamefont {{Montenegro-Montes}},\ and\ \citenamefont {{Zimmerer}}}]{2024arXiv240218645M}%
  \BibitemOpen
  \bibfield  {author} {\bibinfo {author} {\bibfnamefont {T.}~\bibnamefont {{Mroczkowski}}}, \bibinfo {author} {\bibfnamefont {P.~A.}\ \bibnamefont {{Gallardo}}}, \bibinfo {author} {\bibfnamefont {M.}~\bibnamefont {{Timpe}}}, \bibinfo {author} {\bibfnamefont {A.}~\bibnamefont {{Kiselev}}}, \bibinfo {author} {\bibfnamefont {M.}~\bibnamefont {{Groh}}}, \bibinfo {author} {\bibfnamefont {H.}~\bibnamefont {{Kaercher}}}, \bibinfo {author} {\bibfnamefont {M.}~\bibnamefont {{Reichert}}}, \bibinfo {author} {\bibfnamefont {C.}~\bibnamefont {{Cicone}}}, \bibinfo {author} {\bibfnamefont {R.}~\bibnamefont {{Puddu}}}, \bibinfo {author} {\bibfnamefont {P.}~\bibnamefont {{Dubois-dit-Bonclaude}}}, \bibinfo {author} {\bibfnamefont {D.}~\bibnamefont {{Bok}}}, \bibinfo {author} {\bibfnamefont {E.}~\bibnamefont {{Dahl}}}, \bibinfo {author} {\bibfnamefont {M.}~\bibnamefont {{Macintosh}}}, \bibinfo {author} {\bibfnamefont {S.}~\bibnamefont {{Dicker}}}, \bibinfo {author} {\bibfnamefont {I.}~\bibnamefont {{Viole}}}, \bibinfo {author}
  {\bibfnamefont {S.}~\bibnamefont {{Sartori}}}, \bibinfo {author} {\bibfnamefont {G.}~\bibnamefont {{Andr{\'e}s Valenzuela Venegas}}}, \bibinfo {author} {\bibfnamefont {M.}~\bibnamefont {{Zeyringer}}}, \bibinfo {author} {\bibfnamefont {M.}~\bibnamefont {{Niemack}}}, \bibinfo {author} {\bibfnamefont {S.}~\bibnamefont {{Poppi}}}, \bibinfo {author} {\bibfnamefont {R.}~\bibnamefont {{Olguin}}}, \bibinfo {author} {\bibfnamefont {E.}~\bibnamefont {{Hatziminaoglou}}}, \bibinfo {author} {\bibfnamefont {C.}~\bibnamefont {{De Breuck}}}, \bibinfo {author} {\bibfnamefont {P.}~\bibnamefont {{Klaassen}}}, \bibinfo {author} {\bibfnamefont {F.~M.}\ \bibnamefont {{Montenegro-Montes}}}, \ and\ \bibinfo {author} {\bibfnamefont {T.}~\bibnamefont {{Zimmerer}}},\ }\href {\doibase 10.1051/0004-6361/202449786} {\bibfield  {journal} {\bibinfo  {journal} {\aap}\ }\textbf {\bibinfo {volume} {694}},\ \bibinfo {eid} {A142} (\bibinfo {year} {2025})},\ \Eprint {http://arxiv.org/abs/2402.18645} {arXiv:2402.18645 [astro-ph.IM]} \BibitemShut
  {NoStop}%
\bibitem [{\citenamefont {{van Marrewijk}}\ \emph {et~al.}(2024)\citenamefont {{van Marrewijk}}, \citenamefont {{Morris}}, \citenamefont {{Mroczkowski}}, \citenamefont {{Cicone}}, \citenamefont {{Dicker}}, \citenamefont {{Di Mascolo}}, \citenamefont {{Haridas}}, \citenamefont {{Orlowski-Scherer}}, \citenamefont {{Rasia}}, \citenamefont {{Romero}},\ and\ \citenamefont {{W{\"u}rzinger}}}]{2024OJAp....7E.118V}%
  \BibitemOpen
  \bibfield  {author} {\bibinfo {author} {\bibfnamefont {J.}~\bibnamefont {{van Marrewijk}}}, \bibinfo {author} {\bibfnamefont {T.~W.}\ \bibnamefont {{Morris}}}, \bibinfo {author} {\bibfnamefont {T.}~\bibnamefont {{Mroczkowski}}}, \bibinfo {author} {\bibfnamefont {C.}~\bibnamefont {{Cicone}}}, \bibinfo {author} {\bibfnamefont {S.}~\bibnamefont {{Dicker}}}, \bibinfo {author} {\bibfnamefont {L.}~\bibnamefont {{Di Mascolo}}}, \bibinfo {author} {\bibfnamefont {S.~K.}\ \bibnamefont {{Haridas}}}, \bibinfo {author} {\bibfnamefont {J.}~\bibnamefont {{Orlowski-Scherer}}}, \bibinfo {author} {\bibfnamefont {E.}~\bibnamefont {{Rasia}}}, \bibinfo {author} {\bibfnamefont {C.}~\bibnamefont {{Romero}}}, \ and\ \bibinfo {author} {\bibfnamefont {J.}~\bibnamefont {{W{\"u}rzinger}}},\ }\href {\doibase 10.33232/001c.127571} {\bibfield  {journal} {\bibinfo  {journal} {The Open Journal of Astrophysics}\ }\textbf {\bibinfo {volume} {7}},\ \bibinfo {eid} {118} (\bibinfo {year} {2024})},\ \Eprint {http://arxiv.org/abs/2402.10731}
  {arXiv:2402.10731 [astro-ph.IM]} \BibitemShut {NoStop}%
\bibitem [{\citenamefont {{De Moortel}}\ and\ \citenamefont {{Browning}}(2015)}]{2015RSPTA.37340269D}%
  \BibitemOpen
  \bibfield  {author} {\bibinfo {author} {\bibfnamefont {I.}~\bibnamefont {{De Moortel}}}\ and\ \bibinfo {author} {\bibfnamefont {P.}~\bibnamefont {{Browning}}},\ }\href {\doibase 10.1098/rsta.2014.0269} {\bibfield  {journal} {\bibinfo  {journal} {Philosophical Transactions of the Royal Society of London Series A}\ }\textbf {\bibinfo {volume} {373}},\ \bibinfo {pages} {20140269} (\bibinfo {year} {2015})},\ \Eprint {http://arxiv.org/abs/1510.00977} {arXiv:1510.00977 [astro-ph.SR]} \BibitemShut {NoStop}%
\bibitem [{\citenamefont {{Benz}}(2017)}]{2017LRSP...14....2B}%
  \BibitemOpen
  \bibfield  {author} {\bibinfo {author} {\bibfnamefont {A.~O.}\ \bibnamefont {{Benz}}},\ }\href {\doibase 10.1007/s41116-016-0004-3} {\bibfield  {journal} {\bibinfo  {journal} {Living Reviews in Solar Physics}\ }\textbf {\bibinfo {volume} {14}},\ \bibinfo {eid} {2} (\bibinfo {year} {2017})}\BibitemShut {NoStop}%
\bibitem [{\citenamefont {{Krucker}}\ \emph {et~al.}(1997)\citenamefont {{Krucker}}, \citenamefont {{Benz}}, \citenamefont {{Bastian}},\ and\ \citenamefont {{Acton}}}]{1997ApJ...488..499K}%
  \BibitemOpen
  \bibfield  {author} {\bibinfo {author} {\bibfnamefont {S.}~\bibnamefont {{Krucker}}}, \bibinfo {author} {\bibfnamefont {A.~O.}\ \bibnamefont {{Benz}}}, \bibinfo {author} {\bibfnamefont {T.~S.}\ \bibnamefont {{Bastian}}}, \ and\ \bibinfo {author} {\bibfnamefont {L.~W.}\ \bibnamefont {{Acton}}},\ }\href {\doibase 10.1086/304686} {\bibfield  {journal} {\bibinfo  {journal} {\apj}\ }\textbf {\bibinfo {volume} {488}},\ \bibinfo {pages} {499} (\bibinfo {year} {1997})}\BibitemShut {NoStop}%
\bibitem [{\citenamefont {{Benz}}\ and\ \citenamefont {{Krucker}}(1998)}]{1998SoPh..182..349B}%
  \BibitemOpen
  \bibfield  {author} {\bibinfo {author} {\bibfnamefont {A.~O.}\ \bibnamefont {{Benz}}}\ and\ \bibinfo {author} {\bibfnamefont {S.}~\bibnamefont {{Krucker}}},\ }\href {\doibase 10.1023/A:1005046620684} {\bibfield  {journal} {\bibinfo  {journal} {\solphys}\ }\textbf {\bibinfo {volume} {182}},\ \bibinfo {pages} {349} (\bibinfo {year} {1998})}\BibitemShut {NoStop}%
\bibitem [{\citenamefont {{Berghmans}}\ \emph {et~al.}(1998)\citenamefont {{Berghmans}}, \citenamefont {{Clette}},\ and\ \citenamefont {{Moses}}}]{1998A&A...336.1039B}%
  \BibitemOpen
  \bibfield  {author} {\bibinfo {author} {\bibfnamefont {D.}~\bibnamefont {{Berghmans}}}, \bibinfo {author} {\bibfnamefont {F.}~\bibnamefont {{Clette}}}, \ and\ \bibinfo {author} {\bibfnamefont {D.}~\bibnamefont {{Moses}}},\ }\href@noop {} {\bibfield  {journal} {\bibinfo  {journal} {\aap}\ }\textbf {\bibinfo {volume} {336}},\ \bibinfo {pages} {1039} (\bibinfo {year} {1998})}\BibitemShut {NoStop}%
\bibitem [{\citenamefont {{R{\'e}gnier}}\ and\ \citenamefont {{Canfield}}(2006)}]{2006A&A...451..319R}%
  \BibitemOpen
  \bibfield  {author} {\bibinfo {author} {\bibfnamefont {S.}~\bibnamefont {{R{\'e}gnier}}}\ and\ \bibinfo {author} {\bibfnamefont {R.~C.}\ \bibnamefont {{Canfield}}},\ }\href {\doibase 10.1051/0004-6361:20054171} {\bibfield  {journal} {\bibinfo  {journal} {\aap}\ }\textbf {\bibinfo {volume} {451}},\ \bibinfo {pages} {319} (\bibinfo {year} {2006})}\BibitemShut {NoStop}%
\bibitem [{\citenamefont {{Vial}}\ and\ \citenamefont {{Engvold}}(2015)}]{2015ASSL..415.....V}%
  \BibitemOpen
  \bibinfo {editor} {\bibfnamefont {J.-C.}\ \bibnamefont {{Vial}}}\ and\ \bibinfo {editor} {\bibfnamefont {O.}~\bibnamefont {{Engvold}}},\ eds.,\ \href {\doibase 10.1007/978-3-319-10416-4} {\emph {\bibinfo {title} {Solar Prominences}}},\ \bibinfo {series} {Astrophysics and Space Science Library}, Vol.\ \bibinfo {volume} {415}\ (\bibinfo {year} {2015})\BibitemShut {NoStop}%
\bibitem [{\citenamefont {{Mackay}}\ \emph {et~al.}(2010)\citenamefont {{Mackay}}, \citenamefont {{Karpen}}, \citenamefont {{Ballester}}, \citenamefont {{Schmieder}},\ and\ \citenamefont {{Aulanier}}}]{2010SSRv..151..333M}%
  \BibitemOpen
  \bibfield  {author} {\bibinfo {author} {\bibfnamefont {D.~H.}\ \bibnamefont {{Mackay}}}, \bibinfo {author} {\bibfnamefont {J.~T.}\ \bibnamefont {{Karpen}}}, \bibinfo {author} {\bibfnamefont {J.~L.}\ \bibnamefont {{Ballester}}}, \bibinfo {author} {\bibfnamefont {B.}~\bibnamefont {{Schmieder}}}, \ and\ \bibinfo {author} {\bibfnamefont {G.}~\bibnamefont {{Aulanier}}},\ }\href {\doibase 10.1007/s11214-010-9628-0} {\bibfield  {journal} {\bibinfo  {journal} {\ssr}\ }\textbf {\bibinfo {volume} {151}},\ \bibinfo {pages} {333} (\bibinfo {year} {2010})},\ \Eprint {http://arxiv.org/abs/1001.1635} {arXiv:1001.1635 [astro-ph.SR]} \BibitemShut {NoStop}%
\bibitem [{\citenamefont {{Morris}}\ \emph {et~al.}(2022)\citenamefont {{Morris}}, \citenamefont {{Bustos}}, \citenamefont {{Calabrese}}, \citenamefont {{Choi}}, \citenamefont {{Duivenvoorden}}, \citenamefont {{Dunkley}}, \citenamefont {{D{\"u}nner}}, \citenamefont {{Gallardo}}, \citenamefont {{Hasselfield}}, \citenamefont {{Hincks}}, \citenamefont {{Mroczkowski}}, \citenamefont {{Naess}}, \citenamefont {{Niemack}}, \citenamefont {{Page}}, \citenamefont {{Partridge}}, \citenamefont {{Salatino}}, \citenamefont {{Staggs}}, \citenamefont {{Treu}}, \citenamefont {{Wollack}},\ and\ \citenamefont {{Xu}}}]{2022PhRvD.105d2004M}%
  \BibitemOpen
  \bibfield  {author} {\bibinfo {author} {\bibfnamefont {T.~W.}\ \bibnamefont {{Morris}}}, \bibinfo {author} {\bibfnamefont {R.}~\bibnamefont {{Bustos}}}, \bibinfo {author} {\bibfnamefont {E.}~\bibnamefont {{Calabrese}}}, \bibinfo {author} {\bibfnamefont {S.~K.}\ \bibnamefont {{Choi}}}, \bibinfo {author} {\bibfnamefont {A.~J.}\ \bibnamefont {{Duivenvoorden}}}, \bibinfo {author} {\bibfnamefont {J.}~\bibnamefont {{Dunkley}}}, \bibinfo {author} {\bibfnamefont {R.}~\bibnamefont {{D{\"u}nner}}}, \bibinfo {author} {\bibfnamefont {P.~A.}\ \bibnamefont {{Gallardo}}}, \bibinfo {author} {\bibfnamefont {M.}~\bibnamefont {{Hasselfield}}}, \bibinfo {author} {\bibfnamefont {A.~D.}\ \bibnamefont {{Hincks}}}, \bibinfo {author} {\bibfnamefont {T.}~\bibnamefont {{Mroczkowski}}}, \bibinfo {author} {\bibfnamefont {S.}~\bibnamefont {{Naess}}}, \bibinfo {author} {\bibfnamefont {M.~D.}\ \bibnamefont {{Niemack}}}, \bibinfo {author} {\bibfnamefont {L.}~\bibnamefont {{Page}}}, \bibinfo {author} {\bibfnamefont {B.}~\bibnamefont
  {{Partridge}}}, \bibinfo {author} {\bibfnamefont {M.}~\bibnamefont {{Salatino}}}, \bibinfo {author} {\bibfnamefont {S.}~\bibnamefont {{Staggs}}}, \bibinfo {author} {\bibfnamefont {J.}~\bibnamefont {{Treu}}}, \bibinfo {author} {\bibfnamefont {E.~J.}\ \bibnamefont {{Wollack}}}, \ and\ \bibinfo {author} {\bibfnamefont {Z.}~\bibnamefont {{Xu}}},\ }\href {\doibase 10.1103/PhysRevD.105.042004} {\bibfield  {journal} {\bibinfo  {journal} {\prd}\ }\textbf {\bibinfo {volume} {105}},\ \bibinfo {eid} {042004} (\bibinfo {year} {2022})},\ \Eprint {http://arxiv.org/abs/2111.01319} {arXiv:2111.01319 [astro-ph.IM]} \BibitemShut {NoStop}%
\bibitem [{\citenamefont {{Wedemeyer}}\ \emph {et~al.}(2020)\citenamefont {{Wedemeyer}}, \citenamefont {{Szydlarski}}, \citenamefont {{Jafarzadeh}}, \citenamefont {{Eklund}}, \citenamefont {{Guevara Gomez}}, \citenamefont {{Bastian}}, \citenamefont {{Fleck}}, \citenamefont {{de la Cruz Rodriguez}}, \citenamefont {{Rodger}},\ and\ \citenamefont {{Carlsson}}}]{2020A&A...635A..71W}%
  \BibitemOpen
  \bibfield  {author} {\bibinfo {author} {\bibfnamefont {S.}~\bibnamefont {{Wedemeyer}}}, \bibinfo {author} {\bibfnamefont {M.}~\bibnamefont {{Szydlarski}}}, \bibinfo {author} {\bibfnamefont {S.}~\bibnamefont {{Jafarzadeh}}}, \bibinfo {author} {\bibfnamefont {H.}~\bibnamefont {{Eklund}}}, \bibinfo {author} {\bibfnamefont {J.~C.}\ \bibnamefont {{Guevara Gomez}}}, \bibinfo {author} {\bibfnamefont {T.}~\bibnamefont {{Bastian}}}, \bibinfo {author} {\bibfnamefont {B.}~\bibnamefont {{Fleck}}}, \bibinfo {author} {\bibfnamefont {J.}~\bibnamefont {{de la Cruz Rodriguez}}}, \bibinfo {author} {\bibfnamefont {A.}~\bibnamefont {{Rodger}}}, \ and\ \bibinfo {author} {\bibfnamefont {M.}~\bibnamefont {{Carlsson}}},\ }\href {\doibase 10.1051/0004-6361/201937122} {\bibfield  {journal} {\bibinfo  {journal} {\aap}\ }\textbf {\bibinfo {volume} {635}},\ \bibinfo {eid} {A71} (\bibinfo {year} {2020})},\ \Eprint {http://arxiv.org/abs/2001.02185} {arXiv:2001.02185 [astro-ph.SR]} \BibitemShut {NoStop}%
\bibitem [{\citenamefont {{Lemen}}\ \emph {et~al.}(2012)\citenamefont {{Lemen}}, \citenamefont {{Title}}, \citenamefont {{Akin}}, \citenamefont {{Boerner}}, \citenamefont {{Chou}}, \citenamefont {{Drake}}, \citenamefont {{Duncan}}, \citenamefont {{Edwards}}, \citenamefont {{Friedlaender}}, \citenamefont {{Heyman}}, \citenamefont {{Hurlburt}}, \citenamefont {{Katz}}, \citenamefont {{Kushner}}, \citenamefont {{Levay}}, \citenamefont {{Lindgren}}, \citenamefont {{Mathur}}, \citenamefont {{McFeaters}}, \citenamefont {{Mitchell}}, \citenamefont {{Rehse}}, \citenamefont {{Schrijver}}, \citenamefont {{Springer}}, \citenamefont {{Stern}}, \citenamefont {{Tarbell}}, \citenamefont {{Wuelser}}, \citenamefont {{Wolfson}}, \citenamefont {{Yanari}}, \citenamefont {{Bookbinder}}, \citenamefont {{Cheimets}}, \citenamefont {{Caldwell}}, \citenamefont {{Deluca}}, \citenamefont {{Gates}}, \citenamefont {{Golub}}, \citenamefont {{Park}}, \citenamefont {{Podgorski}}, \citenamefont {{Bush}}, \citenamefont {{Scherrer}},
  \citenamefont {{Gummin}}, \citenamefont {{Smith}}, \citenamefont {{Auker}}, \citenamefont {{Jerram}}, \citenamefont {{Pool}}, \citenamefont {{Soufli}}, \citenamefont {{Windt}}, \citenamefont {{Beardsley}}, \citenamefont {{Clapp}}, \citenamefont {{Lang}},\ and\ \citenamefont {{Waltham}}}]{2012SoPh..275...17L}%
  \BibitemOpen
  \bibfield  {author} {\bibinfo {author} {\bibfnamefont {J.~R.}\ \bibnamefont {{Lemen}}}, \bibinfo {author} {\bibfnamefont {A.~M.}\ \bibnamefont {{Title}}}, \bibinfo {author} {\bibfnamefont {D.~J.}\ \bibnamefont {{Akin}}}, \bibinfo {author} {\bibfnamefont {P.~F.}\ \bibnamefont {{Boerner}}}, \bibinfo {author} {\bibfnamefont {C.}~\bibnamefont {{Chou}}}, \bibinfo {author} {\bibfnamefont {J.~F.}\ \bibnamefont {{Drake}}}, \bibinfo {author} {\bibfnamefont {D.~W.}\ \bibnamefont {{Duncan}}}, \bibinfo {author} {\bibfnamefont {C.~G.}\ \bibnamefont {{Edwards}}}, \bibinfo {author} {\bibfnamefont {F.~M.}\ \bibnamefont {{Friedlaender}}}, \bibinfo {author} {\bibfnamefont {G.~F.}\ \bibnamefont {{Heyman}}}, \bibinfo {author} {\bibfnamefont {N.~E.}\ \bibnamefont {{Hurlburt}}}, \bibinfo {author} {\bibfnamefont {N.~L.}\ \bibnamefont {{Katz}}}, \bibinfo {author} {\bibfnamefont {G.~D.}\ \bibnamefont {{Kushner}}}, \bibinfo {author} {\bibfnamefont {M.}~\bibnamefont {{Levay}}}, \bibinfo {author} {\bibfnamefont {R.~W.}\ \bibnamefont
  {{Lindgren}}}, \bibinfo {author} {\bibfnamefont {D.~P.}\ \bibnamefont {{Mathur}}}, \bibinfo {author} {\bibfnamefont {E.~L.}\ \bibnamefont {{McFeaters}}}, \bibinfo {author} {\bibfnamefont {S.}~\bibnamefont {{Mitchell}}}, \bibinfo {author} {\bibfnamefont {R.~A.}\ \bibnamefont {{Rehse}}}, \bibinfo {author} {\bibfnamefont {C.~J.}\ \bibnamefont {{Schrijver}}}, \bibinfo {author} {\bibfnamefont {L.~A.}\ \bibnamefont {{Springer}}}, \bibinfo {author} {\bibfnamefont {R.~A.}\ \bibnamefont {{Stern}}}, \bibinfo {author} {\bibfnamefont {T.~D.}\ \bibnamefont {{Tarbell}}}, \bibinfo {author} {\bibfnamefont {J.-P.}\ \bibnamefont {{Wuelser}}}, \bibinfo {author} {\bibfnamefont {C.~J.}\ \bibnamefont {{Wolfson}}}, \bibinfo {author} {\bibfnamefont {C.}~\bibnamefont {{Yanari}}}, \bibinfo {author} {\bibfnamefont {J.~A.}\ \bibnamefont {{Bookbinder}}}, \bibinfo {author} {\bibfnamefont {P.~N.}\ \bibnamefont {{Cheimets}}}, \bibinfo {author} {\bibfnamefont {D.}~\bibnamefont {{Caldwell}}}, \bibinfo {author} {\bibfnamefont {E.~E.}\
  \bibnamefont {{Deluca}}}, \bibinfo {author} {\bibfnamefont {R.}~\bibnamefont {{Gates}}}, \bibinfo {author} {\bibfnamefont {L.}~\bibnamefont {{Golub}}}, \bibinfo {author} {\bibfnamefont {S.}~\bibnamefont {{Park}}}, \bibinfo {author} {\bibfnamefont {W.~A.}\ \bibnamefont {{Podgorski}}}, \bibinfo {author} {\bibfnamefont {R.~I.}\ \bibnamefont {{Bush}}}, \bibinfo {author} {\bibfnamefont {P.~H.}\ \bibnamefont {{Scherrer}}}, \bibinfo {author} {\bibfnamefont {M.~A.}\ \bibnamefont {{Gummin}}}, \bibinfo {author} {\bibfnamefont {P.}~\bibnamefont {{Smith}}}, \bibinfo {author} {\bibfnamefont {G.}~\bibnamefont {{Auker}}}, \bibinfo {author} {\bibfnamefont {P.}~\bibnamefont {{Jerram}}}, \bibinfo {author} {\bibfnamefont {P.}~\bibnamefont {{Pool}}}, \bibinfo {author} {\bibfnamefont {R.}~\bibnamefont {{Soufli}}}, \bibinfo {author} {\bibfnamefont {D.~L.}\ \bibnamefont {{Windt}}}, \bibinfo {author} {\bibfnamefont {S.}~\bibnamefont {{Beardsley}}}, \bibinfo {author} {\bibfnamefont {M.}~\bibnamefont {{Clapp}}}, \bibinfo {author}
  {\bibfnamefont {J.}~\bibnamefont {{Lang}}}, \ and\ \bibinfo {author} {\bibfnamefont {N.}~\bibnamefont {{Waltham}}},\ }\href {\doibase 10.1007/s11207-011-9776-8} {\bibfield  {journal} {\bibinfo  {journal} {\solphys}\ }\textbf {\bibinfo {volume} {275}},\ \bibinfo {pages} {17} (\bibinfo {year} {2012})}\BibitemShut {NoStop}%
\bibitem [{\citenamefont {{Pesnell}}\ \emph {et~al.}(2012)\citenamefont {{Pesnell}}, \citenamefont {{Thompson}},\ and\ \citenamefont {{Chamberlin}}}]{2012SoPh..275....3P}%
  \BibitemOpen
  \bibfield  {author} {\bibinfo {author} {\bibfnamefont {W.~D.}\ \bibnamefont {{Pesnell}}}, \bibinfo {author} {\bibfnamefont {B.~J.}\ \bibnamefont {{Thompson}}}, \ and\ \bibinfo {author} {\bibfnamefont {P.~C.}\ \bibnamefont {{Chamberlin}}},\ }\href {\doibase 10.1007/s11207-011-9841-3} {\bibfield  {journal} {\bibinfo  {journal} {\solphys}\ }\textbf {\bibinfo {volume} {275}},\ \bibinfo {pages} {3} (\bibinfo {year} {2012})}\BibitemShut {NoStop}%
\bibitem [{\citenamefont {{Remijan}}\ \emph {et~al.}(2019)\citenamefont {{Remijan}}, \citenamefont {{Biggs}}, \citenamefont {{Cortes}}, \citenamefont {{Dent}}, \citenamefont {{Di Franceso}}, \citenamefont {{Fomalont}}, \citenamefont {{Hales}}, \citenamefont {{Kameno}}, \citenamefont {{Mason}}, \citenamefont {{Philips}}, \citenamefont {{Saini}}, \citenamefont {{Vila Vilaro}},\ and\ \citenamefont {{Villard}}}]{2019athb.rept.....R}%
  \BibitemOpen
  \bibfield  {author} {\bibinfo {author} {\bibfnamefont {A.}~\bibnamefont {{Remijan}}}, \bibinfo {author} {\bibfnamefont {A.}~\bibnamefont {{Biggs}}}, \bibinfo {author} {\bibfnamefont {P.~A.}\ \bibnamefont {{Cortes}}}, \bibinfo {author} {\bibfnamefont {B.}~\bibnamefont {{Dent}}}, \bibinfo {author} {\bibfnamefont {J.}~\bibnamefont {{Di Franceso}}}, \bibinfo {author} {\bibfnamefont {E.}~\bibnamefont {{Fomalont}}}, \bibinfo {author} {\bibfnamefont {A.}~\bibnamefont {{Hales}}}, \bibinfo {author} {\bibfnamefont {S.}~\bibnamefont {{Kameno}}}, \bibinfo {author} {\bibfnamefont {B.}~\bibnamefont {{Mason}}}, \bibinfo {author} {\bibfnamefont {N.}~\bibnamefont {{Philips}}}, \bibinfo {author} {\bibfnamefont {K.}~\bibnamefont {{Saini}}}, \bibinfo {author} {\bibfnamefont {B.}~\bibnamefont {{Vila Vilaro}}}, \ and\ \bibinfo {author} {\bibfnamefont {E.}~\bibnamefont {{Villard}}},\ }\href {\doibase 10.5281/zenodo.4511522} {\enquote {\bibinfo {title} {{ALMA Technical Handbook,ALMA Doc. 7.3, ver. 1.1}},}\ }\bibinfo {howpublished}
  {2019, ALMA Technical Handbook,ALMA Doc. 7.3, ver. 1.1ISBN 978-3-923524-66-2} (\bibinfo {year} {2019})\BibitemShut {NoStop}%
\bibitem [{\citenamefont {{Loukitcheva}}\ \emph {et~al.}(2004)\citenamefont {{Loukitcheva}}, \citenamefont {{Solanki}}, \citenamefont {{Carlsson}},\ and\ \citenamefont {{Stein}}}]{2004A&A...419..747L}%
  \BibitemOpen
  \bibfield  {author} {\bibinfo {author} {\bibfnamefont {M.}~\bibnamefont {{Loukitcheva}}}, \bibinfo {author} {\bibfnamefont {S.~K.}\ \bibnamefont {{Solanki}}}, \bibinfo {author} {\bibfnamefont {M.}~\bibnamefont {{Carlsson}}}, \ and\ \bibinfo {author} {\bibfnamefont {R.~F.}\ \bibnamefont {{Stein}}},\ }\href {\doibase 10.1051/0004-6361:20034159} {\bibfield  {journal} {\bibinfo  {journal} {\aap}\ }\textbf {\bibinfo {volume} {419}},\ \bibinfo {pages} {747} (\bibinfo {year} {2004})}\BibitemShut {NoStop}%
\bibitem [{\citenamefont {{van Kampen}}(2024)}]{AtLAST_memo_4}%
  \BibitemOpen
  \bibfield  {author} {\bibinfo {author} {\bibfnamefont {E.}~\bibnamefont {{van Kampen}}},\ }\href@noop {} {\bibfield  {journal} {\bibinfo  {journal} {AtLAST Memo Series}\ }\textbf {\bibinfo {volume} {4}} (\bibinfo {year} {2024})},\ \bibinfo {note} {\url{https://atlast-telescope.org/memo-series/memo-public/instrumentationwgmemo4_29feb2024.pdf}}\BibitemShut {NoStop}%
\bibitem [{\citenamefont {{Gallardo}}\ \emph {et~al.}(2024)\citenamefont {{Gallardo}}, \citenamefont {{Puddu}}, \citenamefont {{Mroczkowski}}, \citenamefont {{Timpe}}, \citenamefont {{Dubois-dit-Bonclaude}}, \citenamefont {{Groh}}, \citenamefont {{Reichert}}, \citenamefont {{Cicone}},\ and\ \citenamefont {{Kaercher}}}]{2024SPIE13094E..28G}%
  \BibitemOpen
  \bibfield  {author} {\bibinfo {author} {\bibfnamefont {P.~A.}\ \bibnamefont {{Gallardo}}}, \bibinfo {author} {\bibfnamefont {R.}~\bibnamefont {{Puddu}}}, \bibinfo {author} {\bibfnamefont {T.}~\bibnamefont {{Mroczkowski}}}, \bibinfo {author} {\bibfnamefont {M.}~\bibnamefont {{Timpe}}}, \bibinfo {author} {\bibfnamefont {P.}~\bibnamefont {{Dubois-dit-Bonclaude}}}, \bibinfo {author} {\bibfnamefont {M.}~\bibnamefont {{Groh}}}, \bibinfo {author} {\bibfnamefont {M.}~\bibnamefont {{Reichert}}}, \bibinfo {author} {\bibfnamefont {C.}~\bibnamefont {{Cicone}}}, \ and\ \bibinfo {author} {\bibfnamefont {H.~J.}\ \bibnamefont {{Kaercher}}},\ }in\ \href {\doibase 10.1117/12.3020272} {\emph {\bibinfo {booktitle} {Ground-based and Airborne Telescopes X}}},\ \bibinfo {series} {Society of Photo-Optical Instrumentation Engineers (SPIE) Conference Series}, Vol.\ \bibinfo {volume} {13094},\ \bibinfo {editor} {edited by\ \bibinfo {editor} {\bibfnamefont {H.~K.}\ \bibnamefont {{Marshall}}}, \bibinfo {editor} {\bibfnamefont
  {J.}~\bibnamefont {{Spyromilio}}}, \ and\ \bibinfo {editor} {\bibfnamefont {T.}~\bibnamefont {{Usuda}}}}\ (\bibinfo {year} {2024})\ p.\ \bibinfo {pages} {1309428},\ \Eprint {http://arxiv.org/abs/2406.11502} {arXiv:2406.11502 [astro-ph.IM]} \BibitemShut {NoStop}%
\bibitem [{\citenamefont {{Puddu}}\ \emph {et~al.}(2024)\citenamefont {{Puddu}}, \citenamefont {{Gallardo}}, \citenamefont {{Mroczkowski}}, \citenamefont {{Dubois-dit-Bonclaude}}, \citenamefont {{Groh}}, \citenamefont {{Kiselev}}, \citenamefont {{Reichert}}, \citenamefont {{Timpe}}, \citenamefont {{Cicone}}, \citenamefont {{Kaercher}},\ and\ \citenamefont {{D{\"u}nner}}}]{2024SPIE13094E..4SP}%
  \BibitemOpen
  \bibfield  {author} {\bibinfo {author} {\bibfnamefont {R.}~\bibnamefont {{Puddu}}}, \bibinfo {author} {\bibfnamefont {P.~A.}\ \bibnamefont {{Gallardo}}}, \bibinfo {author} {\bibfnamefont {T.}~\bibnamefont {{Mroczkowski}}}, \bibinfo {author} {\bibfnamefont {P.}~\bibnamefont {{Dubois-dit-Bonclaude}}}, \bibinfo {author} {\bibfnamefont {M.}~\bibnamefont {{Groh}}}, \bibinfo {author} {\bibfnamefont {A.}~\bibnamefont {{Kiselev}}}, \bibinfo {author} {\bibfnamefont {M.}~\bibnamefont {{Reichert}}}, \bibinfo {author} {\bibfnamefont {M.}~\bibnamefont {{Timpe}}}, \bibinfo {author} {\bibfnamefont {C.}~\bibnamefont {{Cicone}}}, \bibinfo {author} {\bibfnamefont {H.~J.}\ \bibnamefont {{Kaercher}}}, \ and\ \bibinfo {author} {\bibfnamefont {R.}~\bibnamefont {{D{\"u}nner}}},\ }in\ \href {\doibase 10.1117/12.3020773} {\emph {\bibinfo {booktitle} {Ground-based and Airborne Telescopes X}}},\ \bibinfo {series} {Society of Photo-Optical Instrumentation Engineers (SPIE) Conference Series}, Vol.\ \bibinfo {volume} {13094},\ \bibinfo
  {editor} {edited by\ \bibinfo {editor} {\bibfnamefont {H.~K.}\ \bibnamefont {{Marshall}}}, \bibinfo {editor} {\bibfnamefont {J.}~\bibnamefont {{Spyromilio}}}, \ and\ \bibinfo {editor} {\bibfnamefont {T.}~\bibnamefont {{Usuda}}}}\ (\bibinfo {year} {2024})\ p.\ \bibinfo {pages} {130944S},\ \Eprint {http://arxiv.org/abs/2406.16602} {arXiv:2406.16602 [astro-ph.IM]} \BibitemShut {NoStop}%
\bibitem [{\citenamefont {{Dicker}}\ \emph {et~al.}(2014)\citenamefont {{Dicker}}, \citenamefont {{Ade}}, \citenamefont {{Aguirre}}, \citenamefont {{Brevik}}, \citenamefont {{Cho}}, \citenamefont {{Datta}}, \citenamefont {{Devlin}}, \citenamefont {{Dober}}, \citenamefont {{Egan}}, \citenamefont {{Ford}}, \citenamefont {{Ford}}, \citenamefont {{Hilton}}, \citenamefont {{Irwin}}, \citenamefont {{Mason}}, \citenamefont {{Marganian}}, \citenamefont {{Mello}}, \citenamefont {{McMahon}}, \citenamefont {{Mroczkowski}}, \citenamefont {{Rosenman}}, \citenamefont {{Tucker}}, \citenamefont {{Vale}}, \citenamefont {{White}}, \citenamefont {{Whitehead}},\ and\ \citenamefont {{Young}}}]{2014JLTP..176..808D}%
  \BibitemOpen
  \bibfield  {author} {\bibinfo {author} {\bibfnamefont {S.~R.}\ \bibnamefont {{Dicker}}}, \bibinfo {author} {\bibfnamefont {P.~A.~R.}\ \bibnamefont {{Ade}}}, \bibinfo {author} {\bibfnamefont {J.}~\bibnamefont {{Aguirre}}}, \bibinfo {author} {\bibfnamefont {J.~A.}\ \bibnamefont {{Brevik}}}, \bibinfo {author} {\bibfnamefont {H.~M.}\ \bibnamefont {{Cho}}}, \bibinfo {author} {\bibfnamefont {R.}~\bibnamefont {{Datta}}}, \bibinfo {author} {\bibfnamefont {M.~J.}\ \bibnamefont {{Devlin}}}, \bibinfo {author} {\bibfnamefont {B.}~\bibnamefont {{Dober}}}, \bibinfo {author} {\bibfnamefont {D.}~\bibnamefont {{Egan}}}, \bibinfo {author} {\bibfnamefont {J.}~\bibnamefont {{Ford}}}, \bibinfo {author} {\bibfnamefont {P.}~\bibnamefont {{Ford}}}, \bibinfo {author} {\bibfnamefont {G.}~\bibnamefont {{Hilton}}}, \bibinfo {author} {\bibfnamefont {K.~D.}\ \bibnamefont {{Irwin}}}, \bibinfo {author} {\bibfnamefont {B.~S.}\ \bibnamefont {{Mason}}}, \bibinfo {author} {\bibfnamefont {P.}~\bibnamefont {{Marganian}}}, \bibinfo {author}
  {\bibfnamefont {M.}~\bibnamefont {{Mello}}}, \bibinfo {author} {\bibfnamefont {J.~J.}\ \bibnamefont {{McMahon}}}, \bibinfo {author} {\bibfnamefont {T.}~\bibnamefont {{Mroczkowski}}}, \bibinfo {author} {\bibfnamefont {M.}~\bibnamefont {{Rosenman}}}, \bibinfo {author} {\bibfnamefont {C.}~\bibnamefont {{Tucker}}}, \bibinfo {author} {\bibfnamefont {L.}~\bibnamefont {{Vale}}}, \bibinfo {author} {\bibfnamefont {S.}~\bibnamefont {{White}}}, \bibinfo {author} {\bibfnamefont {M.}~\bibnamefont {{Whitehead}}}, \ and\ \bibinfo {author} {\bibfnamefont {A.~H.}\ \bibnamefont {{Young}}},\ }\href {\doibase 10.1007/s10909-013-1070-8} {\bibfield  {journal} {\bibinfo  {journal} {Journal of Low Temperature Physics}\ }\textbf {\bibinfo {volume} {176}},\ \bibinfo {pages} {808} (\bibinfo {year} {2014})}\BibitemShut {NoStop}%
\bibitem [{\citenamefont {{Griffin}}\ \emph {et~al.}(2002)\citenamefont {{Griffin}}, \citenamefont {{Bock}},\ and\ \citenamefont {{Gear}}}]{2002ApOpt..41.6543G}%
  \BibitemOpen
  \bibfield  {author} {\bibinfo {author} {\bibfnamefont {M.~J.}\ \bibnamefont {{Griffin}}}, \bibinfo {author} {\bibfnamefont {J.~J.}\ \bibnamefont {{Bock}}}, \ and\ \bibinfo {author} {\bibfnamefont {W.~K.}\ \bibnamefont {{Gear}}},\ }\href {\doibase 10.1364/AO.41.006543} {\bibfield  {journal} {\bibinfo  {journal} {\ao}\ }\textbf {\bibinfo {volume} {41}},\ \bibinfo {pages} {6543} (\bibinfo {year} {2002})},\ \Eprint {http://arxiv.org/abs/astro-ph/0205264} {arXiv:astro-ph/0205264 [astro-ph]} \BibitemShut {NoStop}%
\bibitem [{\citenamefont {{Mason}}(2003)}]{daisy_gbt}%
  \BibitemOpen
  \bibfield  {author} {\bibinfo {author} {\bibfnamefont {B.}~\bibnamefont {{Mason}}},\ }\href {{https://www.gb.nrao.edu/ptcs/ptcspn/ptcspn33/ptcspn33.pdf}} {\bibfield  {journal} {\bibinfo  {journal} {{Precision Telescope Control System PTCS Project Note 33.1}}\ } (\bibinfo {year} {2003})}\BibitemShut {NoStop}%
\bibitem [{\citenamefont {{Korngut}}\ \emph {et~al.}(2011)\citenamefont {{Korngut}}, \citenamefont {{Dicker}}, \citenamefont {{Reese}}, \citenamefont {{Mason}}, \citenamefont {{Devlin}}, \citenamefont {{Mroczkowski}}, \citenamefont {{Sarazin}}, \citenamefont {{Sun}},\ and\ \citenamefont {{Sievers}}}]{2011ApJ...734...10K}%
  \BibitemOpen
  \bibfield  {author} {\bibinfo {author} {\bibfnamefont {P.~M.}\ \bibnamefont {{Korngut}}}, \bibinfo {author} {\bibfnamefont {S.~R.}\ \bibnamefont {{Dicker}}}, \bibinfo {author} {\bibfnamefont {E.~D.}\ \bibnamefont {{Reese}}}, \bibinfo {author} {\bibfnamefont {B.~S.}\ \bibnamefont {{Mason}}}, \bibinfo {author} {\bibfnamefont {M.~J.}\ \bibnamefont {{Devlin}}}, \bibinfo {author} {\bibfnamefont {T.}~\bibnamefont {{Mroczkowski}}}, \bibinfo {author} {\bibfnamefont {C.~L.}\ \bibnamefont {{Sarazin}}}, \bibinfo {author} {\bibfnamefont {M.}~\bibnamefont {{Sun}}}, \ and\ \bibinfo {author} {\bibfnamefont {J.}~\bibnamefont {{Sievers}}},\ }\href {\doibase 10.1088/0004-637X/734/1/10} {\bibfield  {journal} {\bibinfo  {journal} {\apj}\ }\textbf {\bibinfo {volume} {734}},\ \bibinfo {eid} {10} (\bibinfo {year} {2011})},\ \Eprint {http://arxiv.org/abs/1010.5494} {arXiv:1010.5494 [astro-ph.CO]} \BibitemShut {NoStop}%
\bibitem [{\citenamefont {{Gun{\'a}r}}\ \emph {et~al.}(2016)\citenamefont {{Gun{\'a}r}}, \citenamefont {{Heinzel}}, \citenamefont {{Mackay}},\ and\ \citenamefont {{Anzer}}}]{2016ApJ...833..141G}%
  \BibitemOpen
  \bibfield  {author} {\bibinfo {author} {\bibfnamefont {S.}~\bibnamefont {{Gun{\'a}r}}}, \bibinfo {author} {\bibfnamefont {P.}~\bibnamefont {{Heinzel}}}, \bibinfo {author} {\bibfnamefont {D.~H.}\ \bibnamefont {{Mackay}}}, \ and\ \bibinfo {author} {\bibfnamefont {U.}~\bibnamefont {{Anzer}}},\ }\href {\doibase 10.3847/1538-4357/833/2/141} {\bibfield  {journal} {\bibinfo  {journal} {\apj}\ }\textbf {\bibinfo {volume} {833}},\ \bibinfo {eid} {141} (\bibinfo {year} {2016})}\BibitemShut {NoStop}%
\bibitem [{\citenamefont {{Gun{\'a}r}}\ \emph {et~al.}(2018)\citenamefont {{Gun{\'a}r}}, \citenamefont {{Heinzel}}, \citenamefont {{Anzer}},\ and\ \citenamefont {{Mackay}}}]{2018ApJ...853...21G}%
  \BibitemOpen
  \bibfield  {author} {\bibinfo {author} {\bibfnamefont {S.}~\bibnamefont {{Gun{\'a}r}}}, \bibinfo {author} {\bibfnamefont {P.}~\bibnamefont {{Heinzel}}}, \bibinfo {author} {\bibfnamefont {U.}~\bibnamefont {{Anzer}}}, \ and\ \bibinfo {author} {\bibfnamefont {D.~H.}\ \bibnamefont {{Mackay}}},\ }\href {\doibase 10.3847/1538-4357/aaa001} {\bibfield  {journal} {\bibinfo  {journal} {\apj}\ }\textbf {\bibinfo {volume} {853}},\ \bibinfo {eid} {21} (\bibinfo {year} {2018})}\BibitemShut {NoStop}%
\bibitem [{\citenamefont {{Parks}}\ and\ \citenamefont {{Winckler}}(1969)}]{1969ApJ...155L.117P}%
  \BibitemOpen
  \bibfield  {author} {\bibinfo {author} {\bibfnamefont {G.~K.}\ \bibnamefont {{Parks}}}\ and\ \bibinfo {author} {\bibfnamefont {J.~R.}\ \bibnamefont {{Winckler}}},\ }\href {\doibase 10.1086/180315} {\bibfield  {journal} {\bibinfo  {journal} {\apjl}\ }\textbf {\bibinfo {volume} {155}},\ \bibinfo {pages} {L117} (\bibinfo {year} {1969})}\BibitemShut {NoStop}%
\bibitem [{\citenamefont {{Phillips}}\ \emph {et~al.}(2015)\citenamefont {{Phillips}}, \citenamefont {{Hills}}, \citenamefont {{Bastian}}, \citenamefont {{Hudson}}, \citenamefont {{Marson}},\ and\ \citenamefont {{Wedemeyer}}}]{2015ASPC..499..347P}%
  \BibitemOpen
  \bibfield  {author} {\bibinfo {author} {\bibfnamefont {N.}~\bibnamefont {{Phillips}}}, \bibinfo {author} {\bibfnamefont {R.}~\bibnamefont {{Hills}}}, \bibinfo {author} {\bibfnamefont {T.}~\bibnamefont {{Bastian}}}, \bibinfo {author} {\bibfnamefont {H.}~\bibnamefont {{Hudson}}}, \bibinfo {author} {\bibfnamefont {R.}~\bibnamefont {{Marson}}}, \ and\ \bibinfo {author} {\bibfnamefont {S.}~\bibnamefont {{Wedemeyer}}},\ }in\ \href {\doibase 10.48550/arXiv.1502.06122} {\emph {\bibinfo {booktitle} {Revolution in Astronomy with ALMA: The Third Year}}},\ \bibinfo {series} {Astronomical Society of the Pacific Conference Series}, Vol.\ \bibinfo {volume} {499},\ \bibinfo {editor} {edited by\ \bibinfo {editor} {\bibfnamefont {D.}~\bibnamefont {{Iono}}}, \bibinfo {editor} {\bibfnamefont {K.}~\bibnamefont {{Tatematsu}}}, \bibinfo {editor} {\bibfnamefont {A.}~\bibnamefont {{Wootten}}}, \ and\ \bibinfo {editor} {\bibfnamefont {L.}~\bibnamefont {{Testi}}}}\ (\bibinfo {year} {2015})\ p.\ \bibinfo {pages} {347},\ \Eprint
  {http://arxiv.org/abs/1502.06122} {arXiv:1502.06122 [astro-ph.IM]} \BibitemShut {NoStop}%
\end{thebibliography}%

\end{document}